\documentclass[sigconf, nonacm]{acmart}
\usepackage[ruled]{algorithm2e} 
\usepackage{bm}
\usepackage{subfig}
\usepackage{xcolor}
\usepackage{hyperref}

\newtheorem{problem}{Problem}
\newtheorem{definition}{Definition}
\newtheorem{theorem}{Theorem}
\newtheorem{lemma}{Lemma}
\newtheorem{corollary}{Corollary}

\begin{document}
\raggedbottom
\title{Target PageRank: Matching Prescribed PageRank Scores via Budgeted Edge Additions}

\author{Parth Gupta$^{*1}$, Abhinav Jain$^{*1}$, Avishek Mukhopadhyay$^{2}$, Amitabha Bagchi$^{1}$, Sayan Ranu$^{1}$ and Abhijnan Chakraborty$^{2}$}

\affiliation{%
  \institution{\rule{0pt}{2.5em}$^{1}$Indian Institute of Technology Delhi\\
  $^{2}$Indian Institute of Technology Kharagpur}}

\begin{teaserfigure}
  \centerline{\small $^{*}$ Both authors contributed equally to this research.}
\end{teaserfigure}








\begin{abstract}
PageRank is widely used to quantify the importance of nodes in directed networks. In applications such as recommendation systems, it is increasingly important that the PageRank distribution provides a fair representation of node exposure. This motivates the problem of making minimal changes to a graph so that the PageRank values of a given set of target nodes $T$ match a desired vector $\boldsymbol{t}$. In this paper, we study this problem under an edge-addition budget $B$, where exactly $B$ edges may be added to the graph. We show that deciding whether a multiset of exactly $B$ added edges can make the target nodes attain their desired absolute PageRank scores is NP-complete. After this, we define the $\ell_1$ \textsc{Target PageRank} problem in which we seek to minimize the $\ell_1$ distance between the target vector and resultant vector. We show that this optimization problem admits no polynomial-time multiplicative constant-factor approximation unless $P=NP$. Motivated by these hardness results, we propose two greedy heuristics for approaching the target PageRank vector. We additionally propose a Random Walk Hungarian (RWH) heuristic that addresses the addition of edges between dissimilar nodes by incorporating a penalty for such edges. Finally, we evaluate the proposed heuristics on several real-world and synthetic datasets and compare their performance in terms of PageRank accuracy, computational efficiency, and the relevance of the added edges.
\end{abstract}

\maketitle

\section{Introduction}

PageRank is a widely used measure of node importance in directed networks, originally introduced for ranking Web pages based on the structure of hyperlinks~\cite{page1998pagerank,brin1998anatomy}. The PageRank of a node depends not only on the number of incoming links but also on the importance of the nodes from which those links originate. This makes PageRank useful beyond Web search, including applications such as recommendation systems. In many such applications, however, the observed PageRank distribution of a network may not correspond to a desired distribution of importance or exposure. This motivates the problem of engineering the link structure of a network to achieve a prescribed PageRank score vector~\cite{dash2022fairir}.

Consider a directed network $G=(V,E)$, a subset of target nodes $T \subseteq V$, and a vector $\boldsymbol{t}$ of desired absolute PageRank scores on $T$. We ask whether there exists a multiset of exactly $B$ edge additions that makes the resulting PageRank scores on $T$ equal to $\boldsymbol{t}$. We call this the \textsc{Target PageRank} problem. Unlike the traditional PageRank manipulation setting, our formulation considers a desired score vector over multiple target nodes and allows the desired PageRank values to be specified explicitly.

We first investigate the computational complexity of this problem. We show that \textsc{Target PageRank} is NP-complete by a reduction from a generalized balanced partition problem. Furthermore, we consider the optimization version, in which the objective is to minimize the $\ell_1$ distance between the resulting and desired PageRank score vectors on $T$. We show that this optimization problem admits no polynomial-time multiplicative constant-factor approximation unless $P=NP$. This worst-case multiplicative hardness does not rule out additive guarantees or effective heuristics on particular instances.

Motivated by these hardness results, we develop three scalable heuristics. The first, \textsc{Greedy 1} (G1), exploits the exposure bias of target nodes and preferentially adds edges from over-exposed nodes to under-exposed nodes, while also considering high- and low-PageRank nodes outside the target set. The second, \textsc{Greedy 2} (G2), explicitly estimates the contribution of candidate edges to the reduction in $\ell_1$ error and greedily selects the most useful edges. Finally, \textsc{Random Walk Hungarian} (RWH) exploits the local structure of the network by identifying clusters of over- and under-exposed nodes using random walks and then uses a minimum-cost matching procedure to select edges while accounting for domain-specific similarity.

We evaluate the proposed heuristics on large real-world recommendation-induced networks constructed from Amazon product-review data, as well as synthetic power-law graphs. Our evaluation considers both the reduction in $\ell_1$ distance to the target score vector and the relevance of the newly added edges. Across our experiments, RWH generally achieves a better trade-off between PageRank error and edge similarity, while G1 is the least computationally expensive heuristic and G2 can provide improved PageRank optimization at a higher computational cost.

Our contributions can be summarized as follows:

\begin{itemize}
\item We formulate the \textsc{Target PageRank} problem, which asks whether exactly $B$ edge additions can make a specified set of nodes attain a desired absolute PageRank score vector.

\item We prove that \textsc{Target PageRank} is \textbf{NP-complete} and that, unless $P=NP$, its $\ell_1$-optimization variant is \textbf{constant-factor inapproximable} under a multiplicative guarantee.

\item We propose three scalable heuristics---\textsc{G1}, \textsc{G2}, and \textsc{RWH}---for approximately solving the optimization problem.

\item We experimentally evaluate these approaches on large real-world and synthetic networks and study their accuracy, scalability, and the relevance of the edges they introduce.

\end{itemize}
\section{Related Work}
A line of work has studied the computational problem of modifying the link structure of a directed network in order to influence PageRank. In particular, Olsen et al.~\cite{olsen2008intractability} studied the computational complexity of increasing the PageRank of a designated target node through the addition of links, establishing hardness results for the resulting optimization problem. Subsequent work by Olsen et al.~\cite{olsen2014approximability} investigated the approximability of PageRank manipulation and established further inapproximability results.

More closely related to our setting is the work of Tsioutsiouliklis et al.~\cite{tsioutsiouliklis2022link}, which studies the effect of edge additions on PageRank fairness. They derive analytical expressions characterizing how adding an edge affects PageRank fairness and use these results to design efficient link-recommendation algorithms for improving fairness. Their objective is to make the allocation of PageRank weight across nodes or groups more equitable, rather than to match a prescribed PageRank score for a specified collection of target nodes. Their algorithms can identify edges whose addition improves a fairness objective and can operate in linear time in the network size ~\cite{tsioutsiouliklis2022link}.

Our \textsc{Target PageRank} problem considers a different objective: given a set of target nodes and an explicitly specified vector of desired absolute PageRank scores, we seek edge additions that make the resulting PageRank vector on those nodes match the prescribed vector. Thus, rather than optimizing a fairness measure or the PageRank of an individual node, our formulation imposes simultaneous constraints on the PageRank values of multiple target nodes. This distinction also leads to a different computational structure: we study the exact feasibility problem and the $\ell_1$ optimization problem induced by the prescribed target vector.
\section{Problem Formulation}

In this section, we introduce the mathematical background for the PageRank algorithm, then define the problem statement and analyze its computability. 

\begin{definition}
\textit{(Weighted Directed Graph): A weighted directed graph 
 $G=(V, E)$ is a tuple of a set of vertices $V$ and a multiset of edges $E$, where each element $e \in E$ is an element of $V \times V$ - $\{(u,u) : u \in V\} $.}
\end{definition}

Let $n=|V|\ge 1$ and let $w_{uv}\in\mathbb{Z}_{\ge 0}$ denote the multiplicity of the edge $(u,v)$, with $w_{uv}=0$ for absent edges and $w_{uu}=0$. The input lists the vertices explicitly and stores one binary-encoded count for each ordered pair with nonzero multiplicity; zero counts are implicit. The out-degree $outdeg_u=\sum_{v\in V}w_{uv}$ counts outgoing edge copies.


We now define PageRank for a fixed rational damping factor $0<\alpha<1$, independent of the input. The experiments use $\alpha=17/20$. At each step, a random surfer follows a uniformly chosen outgoing edge copy with probability $\alpha$ and teleports to a uniformly chosen vertex with probability $1-\alpha$. At a dangling node (a node of out-degree zero), the next vertex is chosen uniformly from all of $V$, including the current vertex. This redistribution is a transition rule and creates no graph edges. Transitions are recomputed after every edge addition: once a dangling node gains an outgoing edge, its uniform sink redistribution disappears and edge-following transitions use its actual outgoing edge multiplicities.

\begin{definition}
\label{def:pagerank}
\textit{(PageRank): Given a graph $G=(V, E)$, its PageRank vector $\boldsymbol{\pi}\in\mathbb{R}^{V}$ satisfies, for every $v\in V$,}

\begin{equation}
\label{eq:pagerank}
\begin{aligned}
\boldsymbol{\pi}_v ={}& \alpha\sum_{\substack{u\in V\\outdeg_u>0}}
\frac{w_{uv}\boldsymbol{\pi}_u}{outdeg_u} \\
&+\frac{\alpha}{n}\sum_{\substack{u\in V\\outdeg_u=0}}\boldsymbol{\pi}_u
+\frac{1-\alpha}{n}.
\end{aligned}
\end{equation}

\begin{equation}
\label{eq:pagerank-normalization}
\sum_{v\in V}\boldsymbol{\pi}_v=1,\qquad
\boldsymbol{\pi}_v\ge 0\quad(v\in V).
\end{equation}
\end{definition}

We study a network design problem in which edge additions are used to approach a target score vector $\boldsymbol{t}$ on a subset $T\subseteq V$. Its entries are desired absolute PageRank scores, not probabilities conditioned on membership in $T$. Each $\boldsymbol{t}_v=a_v/b_v\in[0,1]$ is an exact fraction, with binary-encoded integers $a_v\ge 0$ and $b_v>0$. We do not require $\sum_{v\in T}\boldsymbol{t}_v=1$ and do not renormalize scores on $T$.

Let $E'$ denote the multiset of added edges. We write $E\uplus E'$ for addition of multiplicities: if $(u,v)$ occurs $w_{uv}$ times in $E$ and $x_{uv}$ times in $E'$, it occurs $w_{uv}+x_{uv}$ times in $E\uplus E'$. Thus repeated additions, including copies of existing edges, each consume one unit of budget. The budget $B\in\mathbb{Z}_{\ge 0}$ is encoded in binary. Proposed additions are represented by binary-encoded integer counts satisfying
\begin{equation*}
x_{uv}\ge 0,\qquad x_{uu}=0,\qquad
|E'|=\sum_{u,v\in V}x_{uv}=B.
\end{equation*}
We use $\boldsymbol{\tilde{\pi}}(E')$ to denote the PageRank vector of $G'=(V,E\uplus E')$, with out-degrees and dangling-node transitions recomputed on $G'$. In the algorithms below, $E$ denotes the original edge multiset and $E'$ accumulates additions. These encoding conventions apply to both problems below.

\begin{problem}
\textit{\textsc{Target PageRank} problem:
Given a weighted directed graph $G=(V, E)$, a target set of nodes $T$, a target score vector $\boldsymbol{t}$ and a budget $B$, does there exist an addition multiset $E'$ with $|E'| = B$ such that $e \in \{(u, v) \in V \times V : u \neq v\}$ for all $e \in E'$, and $\boldsymbol{\tilde{\pi}}_v(E') = \boldsymbol{t}_v$ for every $v \in T$?}
\end{problem}

\begin{theorem}
\textit{\textsc{Target PageRank} problem} is NP-Complete.
\end{theorem}

Since \textit{Problem 1} is NP-Complete, we formulate another problem in which the objective is to minimize the distance between the target score vector and the resulting PageRank scores on $T$.
To measure this distance we will use distance function $\ell_1(\boldsymbol{p}, \boldsymbol{q})$ which denotes the $\ell _1$ norm between vectors $\boldsymbol{p}$ and $\boldsymbol{q}$. The subscript $T$ denotes restriction of a vector to the target vertices.

\begin{problem}
\textit{$\ell _1$ \textsc{Target PageRank} problem:
Given a weighted directed graph $G=(V, E)$, a target set of nodes $T$, a target score vector $\boldsymbol{t}$ and a budget $B$, find a multiset of edges $E'$ with $|E'| = B$ such that $e \in \{(u, v) \in V \times V : u \neq v\}$ for all $e \in E'$ and $\ell _1(\boldsymbol{\tilde{\pi}}_T(E'), \boldsymbol{t})$ is minimized.}
\end{problem}

For the optimization problem, feasibility concerns only the addition constraints: an admissible addition multiset of size $B$ exists exactly when $n\ge 2$ or $B=0$, since repeated edges are allowed but self-loops are forbidden. For a feasible instance $I=(G,T,\boldsymbol{t},B)$, write
\begin{equation*}
\begin{aligned}
F_I(E')&=\ell_1(\boldsymbol{\tilde{\pi}}_T(E'),\boldsymbol{t})\\
&=\sum_{v\in T}|\boldsymbol{\tilde{\pi}}_v(E')-\boldsymbol{t}_v|,\\
\operatorname{OPT}(I)&=\min_{E'\ \mathrm{feasible}}F_I(E').
\end{aligned}
\end{equation*}
The minimum ranges over admissible addition multisets of exactly $B$ edges. For $\rho\ge1$, a multiplicative $\rho$-approximation must return a feasible $E'$ satisfying $F_I(E')\le\rho\operatorname{OPT}(I)$ on every feasible instance, including those with $\operatorname{OPT}(I)=0$.

\begin{theorem}
Unless $P=NP$, the \textit{$\ell_1$ \textsc{Target PageRank} problem} admits no deterministic polynomial-time approximation algorithm with any fixed finite multiplicative factor $\rho\ge1$.
\end{theorem}
The same result holds for every $\ell_k$-norm with $1\le k\le\infty$. We prove Theorem 1 and Theorem 2 in Section 10 and Section 11.

\section{Optimal Solution using MILP}
\label{sec:milp}
We formulate the $\ell_1$ \textsc{Target PageRank} problem as an exact mixed-integer linear program (MILP). Let $E'$ denote the multiset of added edges and let $\mathcal{A}=\{(i,j)\in V\times V:i\ne j\}$ be the set of permitted ordered pairs. The input multiplicities $w_{ij}$ count actual edges only. For a budget $B\in\mathbb{Z}_{\ge 0}$, define
\begin{equation*}
d_i=\sum_{j\ne i}w_{ij},\qquad U_i=d_i+B,
\qquad K=\max\{1,\lceil\log_2(B+1)\rceil\}.
\end{equation*}
Throughout this section, $\boldsymbol{\pi}$ denotes the PageRank vector after additions. The formulation uses the following variables:
\begin{itemize}
    \item $\boldsymbol{\pi}_i\in[0,1]$ for $i\in V$ and $\boldsymbol{z}_i\ge 0$ for $i\in T$ are continuous PageRank and absolute-error variables.
    \item $b_{ijr}\in\{0,1\}$, for $(i,j)\in\mathcal{A}$ and $0\le r<K$, encode the number $x_{ij}$ of added copies of $(i,j)$ in binary.
    \item $\delta_i\in\{0,1\}$ indicates whether $i$ is dangling after additions.
    \item $q_i$ is the continuous PageRank mass carried by one outgoing edge copy before damping; it is zero for dangling nodes. The continuous variable $h_i$ represents mass redistributed by a dangling node.
    \item The continuous variables $y_{ijr}$ linearize $b_{ijr}q_i$.
\end{itemize}
The quantities $x_{ij}$, the updated degrees $D_i$, and the flows $F_{ij}$ below are linear expressions in these variables. Vertex sums range over $V$.

The addition counts and exact budget satisfy
\begin{equation}
\label{eq:milp-budget}
\begin{aligned}
x_{ij}&=\sum_{r=0}^{K-1}2^r b_{ijr} &&((i,j)\in\mathcal{A}),\\
\sum_{(i,j)\in\mathcal{A}}x_{ij}&=B,\\
D_i&=d_i+\sum_{j\ne i}x_{ij} &&(i\in V).
\end{aligned}
\end{equation}
These constraints imply $0\le x_{ij}\le B$ and integer multiplicities. Only off-diagonal pairs are indexed, so self-loops cannot be added.

For each $i\in V$, dangling status and the auxiliary bounds are enforced by
\begin{equation}
\label{eq:milp-dangling}
\begin{aligned}
1-\delta_i&\le D_i\le U_i(1-\delta_i),\\
0&\le q_i\le 1-\delta_i,\qquad 0\le h_i\le\delta_i.
\end{aligned}
\end{equation}
Since $D_i$ is a nonnegative integer, these constraints give $\delta_i=1$ exactly when $D_i=0$. The unit bounds are valid because PageRank is at most one and every positive degree is at least one.

For every $(i,j)\in\mathcal{A}$ and $0\le r<K$, impose
\begin{equation}
\label{eq:milp-product}
\begin{aligned}
0\le y_{ijr}&\le q_i,\qquad y_{ijr}\le b_{ijr},\\
y_{ijr}&\ge q_i-(1-b_{ijr}).
\end{aligned}
\end{equation}
When $b_{ijr}=0$, these inequalities force $y_{ijr}=0$; when $b_{ijr}=1$, they force $y_{ijr}=q_i$. Thus the product is represented exactly by linear constraints. Define the flow along actual edge copies by
\begin{equation}
\label{eq:milp-flow}
F_{ij}=w_{ij}q_i+\sum_{r=0}^{K-1}2^r y_{ijr},
\qquad (i,j)\in\mathcal{A}.
\end{equation}

Outgoing mass conservation, incoming PageRank balance, and normalization are
\begin{align}
\boldsymbol{\pi}_i&=\sum_{j\ne i}F_{ij}+h_i &&(i\in V),
\label{eq:milp-outgoing}\\
\boldsymbol{\pi}_j&=\alpha\sum_{i\ne j}F_{ij}
+\frac{\alpha}{n}\sum_i h_i+\frac{1-\alpha}{n} &&(j\in V),
\label{eq:milp-pagerank}\\
\sum_i\boldsymbol{\pi}_i&=1.
\label{eq:milp-normalization}
\end{align}
The dangling term in~\eqref{eq:milp-pagerank} contributes to every vertex, including the dangling vertex itself, without introducing graph edges.

Finally, the objective and error constraints are
\begin{equation}
\label{eq:milp-objective}
\begin{aligned}
\text{minimize}\quad &\sum_{i\in T}\boldsymbol{z}_i,\\
\text{subject to}\quad
&\boldsymbol{z}_i\ge\boldsymbol{\pi}_i-\boldsymbol{t}_i &&(i\in T),\\
&\boldsymbol{z}_i\ge\boldsymbol{t}_i-\boldsymbol{\pi}_i &&(i\in T).
\end{aligned}
\end{equation}
The minimization is subject to all the constraints and variable domains above. At an optimum, $\boldsymbol{z}_i=|\boldsymbol{\pi}_i-\boldsymbol{t}_i|$ for each $i\in T$.

To verify equivalence, first suppose $D_i>0$. Then $\delta_i=h_i=0$, and~\eqref{eq:milp-flow}--\eqref{eq:milp-outgoing} imply $\boldsymbol{\pi}_i=D_iq_i$. Consequently, $F_{ij}=\boldsymbol{\pi}_i(w_{ij}+x_{ij})/D_i$. If $D_i=0$, then $q_i=0$, all outgoing flows vanish, and $h_i=\boldsymbol{\pi}_i$. Substituting these two cases into~\eqref{eq:milp-pagerank} gives Definition~\ref{def:pagerank} on $G'=(V,E\uplus E')$. Conversely, any admissible addition multiset and its PageRank vector satisfy the formulation by choosing the corresponding binary digits and auxiliary variables. Hence the MILP has exactly the intended feasible additions and objective values.

The formulation has $O(n^2K)$ variables and constraints and handles $B=0$ by forcing all addition bits to zero. It provides an exact optimization model without discretizing PageRank. The number of candidate pairs and the integer search can make solving large instances expensive; we therefore develop heuristics for larger graphs.

\section{Greedy 1 (G1)}

First we define exposure-bias of a node and then classify them as over/under-exposed : 
\begin{definition}
\textit{(Exposure-Bias:) Given a weighted directed graph $G = (V, E)$ with PageRank vector $\boldsymbol{\pi}$, a target set of nodes $T$ and a target PageRank vector $\boldsymbol{t}$, the exposure-bias of a node $v \in T$ is defined as $\boldsymbol{eb}_v = \boldsymbol{\pi}_v - \boldsymbol{t}_v$}
\end{definition}

\begin{definition}
\textit{(Over-Exposed Node): A node $v \in T$ is over-exposed if $\boldsymbol{eb}_v > 0.$
}
\end{definition}

\begin{definition}
\textit{(Under-Exposed Node): A node $v \in T$ is under-exposed if $\boldsymbol{eb}_v < 0.$
}
\end{definition}

This algorithm aims to redirect pagerank traffic to nodes with pagerank below the target i.e under-exposed nodes. Adding an edge from a source node to target node reduces the outgoing transition probability of the existing edges and allocates a part of it to the newly added edge. Since we want to increase the PageRank of under-exposed nodes and decrease the PageRank of over-exposed nodes,
selecting over-exposed nodes as candidate sources and under-exposed nodes as candidate destinations can be useful in redirecting pagerank traffic from overexposed nodes to underexposed nodes, and for prioritizing endpoints we use exposure bias as a simple criterion. We also consider high-PageRank sources and low-PageRank destinations outside the target set, allowing changes involving non-target nodes to affect target scores through the network. These choices provide an inexpensive heuristic for selecting edges without evaluating the full PageRank effect of every candidate.  In our heuristics, we have a parameter $k$, which defines the number of edges added in each iteration. We now extend the definition of over-exposed, under-exposed, high-pagerank and low-pagerank nodes with $k$ as a parameter.  \\

 \begin{definition}
\textit{($OVER_{k}$): Let $OVER_{k}$ denote the set of top $k$ overexposed nodes. Then, a node $u \in OVER_{k}$ if there are fewer than ${k}$ nodes $v$ satisfying $eb_v > eb_u$.
}
\end{definition}

\begin{definition}
\textit{($UNDER_{k}$): Let $UNDER_{k}$ denote the set of top $k$ underexposed nodes. Then, a node $u \in UNDER_{k}$ if there are fewer than $k$ nodes $v$ satisfying $eb_v < eb_u$.
}
\end{definition}

\begin{definition}
\textit{($HIGH_{k}$): Let $HIGH_{k}$ denote the set of top $k$ high-PageRank nodes. A node $v \in HIGH_k$ if $v \notin T$ and there are fewer than $k$ nodes $u \notin T$ such that $\boldsymbol{\pi}_u > \boldsymbol{\pi}_v$.
}
\end{definition}

\begin{definition}
\textit{($LOW_{k}$): Let $LOW_{k}$ denote the set of top $k$ low-PageRank nodes. A node $v \in LOW_k$ if $v \notin T$ and there are fewer than $k$ nodes $u \notin T$ such that $\boldsymbol{\pi}_u < \boldsymbol{\pi}_v$.
}
\end{definition}

G1 also has a parameter $ou$ ($\leq k$) which controls the fraction of edges added in each iteration from over-exposed nodes to under-exposed nodes. Similarly, it has a parameter $hl$ which controls the fraction of edges added in each iteration with atleast one endpoint not in target set. In each iteration, we compute $OVER_{ou}$, $UNDER_{ou}$, $HIGH_{hl}$ and $LOW_{hl}$ and add $k$ edges between them as described in the following pseudo-code.

\begin{algorithm}
\caption{Greedy 1 (G1)}
\label{alg:generator}
\SetKwProg{generate}{Function \emph{generate}}{}{end}
\SetKwInOut{Input}{input}\SetKwInOut{Output}{output}
\Input{Graph $G=(V, E)$, PageRank vector $\boldsymbol{\pi}$, target set $T$, target PageRank vector $\boldsymbol{t}$, budget $B$, $k$,$ou$, $hl$}
\Output{Edge multiset $E'$, $|E'| = B$}
\BlankLine

$E'$ $\leftarrow\ \{\}$ \;
\For{$itr \in \{1, 2, \dots , \frac{B}{k}\}$}{
    $eb$ $\leftarrow$ Compute the list of exposure-bias of target nodes\;
    Compute the sets $OVER_{ou}$, $UNDER_{ou}$, $HIGH_{hl}$, $LOW_{hl}$ using $eb$ and $\boldsymbol{\pi}$\;
    $over \leftarrow $ Reverse sorted list of $\{\{eb_i, i\} : i \in OVER_{ou}\}$\;
    $under \leftarrow $ Sorted list of $\{\{eb_i, i\} : i \in UNDER_{ou}\}$\;
    $high \leftarrow $ Reverse sorted list of $\{\{\boldsymbol{\pi}_i, i\} : i \in HIGH_{hl}\}$\;
    $low \leftarrow $ Sorted list of $\{\{\boldsymbol{\pi}_i, i\} : i \in LOW_{hl}\}$\;

    $head \leftarrow $ Append $high$ to $over$\;
    $tail \leftarrow $ Append $low$ to $under$\;
    \For{$i \in \{0, 1, \dots, k-1\}$}{
        $\{val_f, node_f\} = head_i$ \;
        $\{val_t, node_t\} = tail_i$ \;
        $E' = E' \uplus \{node_f \rightarrow node_t\}$ \;
    }
    $G \leftarrow  (V,E \uplus E')$\;
    Recompute $\boldsymbol{\pi}$\ for updated graph $G$
}
\Return E'

\end{algorithm}

\subsection{Time Complexity}
 In each iteration, we compute the list $eb$ which takes $O(|V|)$ time. We can compute each of the lists $over$, $under$, $high$ and $low$ using priority queue which takes $O(k*log|V|)$ time since there are $|V|$ insertions/deletions obtained in a priority queue of size atmost $k$. Hence the time complexity of a single iteration is $O(k*log|V| + |V|)$. Moreover, recomputing the PageRank of the graph after inserting $k$ edges takes $O(k*PR\_UPDATE)$ time, where $O(PR\_UPDATE)$ denotes the time taken for PageRank updation after adding a single edge. Since there are $\frac{B}{k}$ iterations, the final time complexity is $O(B*log|V| + |V|*\frac{B}{k} + B*PR\_UPDATE)$.
\section{Greedy 2 (G2)}

In this algorithm we greedily pick edges based on their individual contribution in decreasing the $\ell_1$-norm between the target PageRank vector and the PageRank vector of graph. In order to evaluate  the individual contribution of an edge $(u,v)$, we first compute the approximate new PageRank vector of the graph by adding $(u,v)$ to the graph using algorithm for incremental PageRank update given in ~\cite{bahmani2010fast} and then calculate the resulting $\ell_1$-norm. The decrease in $\ell_1$-norm by adding the edge $(u,v)$ is the contribution of this edge, denoted by $\Delta(u,v)$.\\

 Similar to Greedy 1, in each iteration of Greedy 2 we first compute the sets $OVER_{ou}$, $UNDER_{ou}$, $HIGH_{hl}$ and $LOW_{hl}$. We then sample $ck$ edges from $OVER_{ou} \cup HIGH_{hl}$ to $UNDER_{ou} \cup LOW_{hl}$ and greedily choose the best $k$ edges among them, based on their individual contribution in decreasing the $\ell_1$-norm. Here $c$ is a parameter which controls the number of edges to be sampled.

\begin{algorithm}
\caption{Greedy 2 (G2)}
\label{alg:generator}
\SetKwProg{generate}{Function \emph{generate}}{}{end}
\SetKwInOut{Input}{input}\SetKwInOut{Output}{output}
\Input{Graph $G=(V, E)$, PageRank vector $\boldsymbol{\pi}$, target set $T$, target PageRank vector $\boldsymbol{t}$, budget $B$, $k$, $ou$, $hl$, $c$}
\Output{Edge multiset $E'$, $|E'| = B$}
\BlankLine

$E'$ $\leftarrow\ \{\}$ \;
$PQ \leftarrow $ Empty Min-Priority Queue \;

\For{$itr \in \{1, 2, \dots , \frac{B}{k}\}$}{
    $eb$ $\leftarrow$ Compute the list of exposure-bias of nodes\;
    Compute the sets $OVER_{ou}$, $UNDER_{ou}$, $HIGH_{hl}$, $LOW_{hl}$ using $eb$ and $\boldsymbol{\pi}$\;
    $over \leftarrow $  list of $OVER_{ou}$ nodes\;
    $under \leftarrow $ list of $UNDER_{ou}$ nodes\;
    $high \leftarrow $  list of $HIGH_{hl}$ nodes\;
    $low \leftarrow $ list of $LOW_{hl}$ nodes\;

    $head \leftarrow $ Append $high$ to $over$\;
    $tail \leftarrow $ Append $low$ to $under$\;
    \For{$i \in \{0, 1, \dots, ck-1\}$}{
        $u \leftarrow $ Randomly sample a node from $head$\;
        $v \leftarrow $ Randomly sample a node from $tail$\;
        $contrib \leftarrow \Delta(u,v)$ \;
        $PQ.insert(\{contrib, (u,v)\})$\;
        \If{$PQ.size() > k$}{
            $PQ.pop()$ \;
        }
    }
    \For{$i \in \{0, 1, \dots, k-1\}$}{
        $E' = E' \uplus \{PQ.top().second\}$ \;
        $PQ.pop()$ \;
    }
    $G \leftarrow  (V,E \uplus E')$\;
    Recompute $\boldsymbol{\pi}$\ for updated graph $G$
}

\Return E'

\end{algorithm}

\subsection{Time Complexity}
 In each iteration, we compute the list $eb$ which takes $O(|V|)$ time. We can compute each of the lists $over$, $under$, $high$ and $low$ using priority queue which takes $O((ou+hl)*log|V|)$ time since there are $|V|$ insertions/deletions obtained in a priority queue of size atmost $ou+hl$. Let evaluation of contribution of an edge take $O(PR\_UPDATE)$ time. Hence the time complexity of a single iteration is $O((ou+hl)*log|V| + |V| + c*k*PR\_UPDATE)$. Moreover, updating the PageRank of graph after inserting $k$ edges takes $O(k*PR\_UPDATE)$ time. Since there are $\frac{B}{k}$ iterations, the final time complexity is $O(\frac{B}{k}*(ou+hl)*log|V| + |V|*\frac{B}{k} + B*c*PR\_UPDATE + B*PR\_UPDATE)$.
\section{RWH (Random Walk Hungarian)}

In this algorithm we consider the local structure surrounding candidate endpoints to
identify opportunities for redistributing pagerank traffic among target
nodes. A destination $u$ from which random walks frequently visit
under-exposed nodes may direct additional traffic toward several
nodes with PageRank deficits. Similarly, a source $v$ from which
random walks frequently visit over-exposed nodes is a candidate
for diverting traffic away from regions with excess exposure.
An edge from $v$ to $u$ may therefore help redistribute exposure in desirable way
across these regions. This motivates selecting source ($v$) and destination cluster centers ($u$)
according to random-walk visits to over-exposed and under-exposed
nodes, respectively. We now define centrality score of a node, which will help identify these cluster centers.

\begin{definition}
\textit{(Centrality Score:) For a node $v$ and a given set of nodes $N$, simulate $R$ random walks starting from $v$. Each of these random walks is continued until its first reset. Then $CS(v)$ is defined as the total number of visits to nodes in $N$ among the total visits in $R$ walks. }
\end{definition}

Algorithm 3 describes the procedure for computing the $Centrality$  $Score$.

\begin{algorithm}
\caption{Centrality Score}
\label{alg:generator}
\SetKwProg{generate}{Function \emph{generate}}{}{end}
\SetKwInOut{Input}{input}\SetKwInOut{Output}{output}
\SetKwFunction{FMain}{CentralityScore}
\SetKwFor{Loop}{Loop}{}{EndLoop}
\SetKwProg{Fn}{Function}{:}{}
\Input{Graph $G=(V, E)$, start node $u$, teleportation probability $1-\alpha$, target nodes $T$, number of walks $R$, max walk length $maxlen$}
\Output{Centrality Score of $u$ w.r.t nodes $T$}
\BlankLine

\Fn{\FMain{$G$, $u$, $\alpha$, $T$, $R$, $maxlen$}}{
    $score \leftarrow 0$ \;
    \For{$i \in \{1, 2, \dots, R\}$}{
        $cur,\ curlen \leftarrow u,\ 1$\;
        \uIf{$cur \in T$}{
            $score \leftarrow score + 1$ \;
        }
        \While{$curlen < maxlen$}{
            $x \leftarrow$ Generate random number in [0, 1] \;
            \uIf{$x < 1-\alpha$}{
                break \;
            } 
            \eIf{$outdeg_{cur}=0$}{
                $cur \leftarrow$ Uniformly sample a vertex from $V$ \;
            }{
                $cur \leftarrow$ Destination of a uniformly sampled outgoing edge copy of $cur$ \;
            }
            \uIf{$cur \in T$}{
                $score \leftarrow score + 1$ \;
            }
            $curlen \leftarrow curlen + 1$\;
        }
    }
    \Return $score$\;
}
\end{algorithm}


\begin{definition}
    \textit{(Top k Over-Exposed/High-Pagerank Cluster Centers:) Let $CENTER^o_k$ denote the set of top $k$ over-exposed / high-pagerank cluster centers. Then, an over-exposed / high-pagerank node $u \in CENTER^o_k$ if there are fewer than $k$ over-exposed / high-pagerank nodes $v$ satisfying $CS(v) > CS(u)$ w.r.t. the set of nodes $OVER_{ou}$.
}
\end{definition}

\begin{definition}
    \textit{(Top k Under-Exposed / Low-Pagerank Cluster Centers:) Let $CENTER^u_k$ denote the set of top $k$ under-exposed / low-pagerank cluster centers. Then, an under-exposed / low-pagerank node $u \in CENTER^u_k$ if there are fewer than $k$ under-exposed / low-pagerank nodes $v$ satisfying $CS(v) > CS(u)$ w.r.t. the set of nodes $UNDER_{ou}$.
}
\end{definition}

In practice we need to keep in mind certain constraints, like we need to add edges between similar items and so on. In order to take into account domain-specific factors we now construct a complete directed weighted bipartite graph with edges from $CENTER^o_k$  to $CENTER^u_k$  with cost of an edge $e$ as the domain specific factor (which is negative of item similarity in our case) and use Hungarian algorithm ~\cite{kuhn1955hungarian} to find the min-cost perfect matching for the above bipartite graph. The edges in the matching are then added to $G$.
Algorithm 4 describes the pseudo code for RWH heuristic.

\begin{algorithm}
\caption{RWH}
\label{alg:generator}
\SetKwProg{generate}{Function \emph{generate}}{}{end}
\SetKwInOut{Input}{input}\SetKwInOut{Output}{output}
\SetKwFunction{FMain}{ComputeCentralityScore}
\SetKwFor{Loop}{Loop}{}{EndLoop}
\SetKwProg{Fn}{Function}{:}{}
\Input{Graph $G=(V, E)$, PageRank vector $\boldsymbol{\pi}$, target set of nodes $T$, target PageRank vector $\boldsymbol{t}$, budget $B$, $k$, $ou$, $hl$}
\Output{Edge multiset $E'$, $|E'| = B$}
\BlankLine

$E'$ $\leftarrow\ \{\}$ \;
\For{$itr \in \{1, 2, \dots , \frac{B}{k}\}$}{
    $eb$ $\leftarrow$ Compute the list of exposure-bias of nodes\;
    Compute the sets $OVER_{ou}$, $UNDER_{ou}$, $HIGH_{hl}$, $LOW_{hl}$ using $eb$ and $\boldsymbol{\pi}$\;    
    $over \leftarrow $  Reverse Sorted list of $\{\{eb_i, i\} : i \in OVER_{ou}\}$\;
    $under \leftarrow $ Sorted list of $\{\{eb_i, i\} : i \in UNDER_{ou}\}$\;
    $high \leftarrow $  Reverse Sorted list of $\{\{\boldsymbol{\pi}_i, i\} : i \in HIGH_{hl}\}$\;
    $low \leftarrow $ Sorted list of $\{\{\boldsymbol{\pi}_i, i\} : i \in LOW_{hl}\}$\;

    $oh \leftarrow $ Append $high$ to $over$\;
    $ul \leftarrow $ Append $low$ to $under$\;
    $cs_o,cs_u \leftarrow \{\}, \{\}$\;
    \For{$(val,u) \in oh$}{
        $cs_o(u) \leftarrow CentralityScore(G, u, \alpha, over, R, len)$ \;
    }
    \For{$(val,u) \in ul$}{
        $cs_u(u) \leftarrow CentralityScore(G, u, \alpha, under, R, len)$ \;
    }
    $head \leftarrow$ Top $k$ cluster centers in $cs_o$ ($CENTER^o_k$) \;
    $tail \leftarrow$ Top $k$ cluster centers in $cs_u$
    ($CENTER^u_k$)\;
    $w \leftarrow \{\}$\;
    \For{$i \in head$}{
        \For{$j \in tail$}{
        $w(i, j) \leftarrow$ $-Similarity(i,j)$ \;
        }
    }
    $G_1 \leftarrow (head \cup tail, head \times tail)$ and edge weights $w$ \;
    $matching \leftarrow$ Compute min-cost max matching for $G_1$ \;
    \For{$e \in matching$}{
        $E' \leftarrow E' \uplus \{e\}$ \;
    }
    $G \leftarrow  (V,E \uplus E')$\;
    Recompute $\boldsymbol{\pi}$\ for updated graph $G$\;
}
\BlankLine
\Return E'

\end{algorithm}

\subsection{Time Complexity}
In each iteration, the computation of $eb$ takes $O(|V|)$ time. The lists $over, under, high, low$ can be computed in $O((ou+hl)*log|V|)$ time using priority queue. The computation of centrality score in one iteration takes $O(k * R * min(\frac{1}{1 - \alpha}, maxlen))$ on an average provided there is a pre-computed hashmap for the set of nodes $T$. The computation of $head$ and $tail$ can be done in $O(k*log|V|)$ using priority queue. We assume that the evaluation of domain-oriented factor of an edge takes $O(DOF)$ time. Then we can construct the bipartite graph in $O(k^2 * DOF)$ time. Finally, a min cost max matching can be computed in $O(k^3)$ time for the constructed graph. Further, recomputing the PageRank of the graph after inserting $k$ edges takes $O(k*PR\_UPDATE)$ time, where $O(PR\_UPDATE)$ denotes the time taken for PageRank updation after adding a single edge. So, the time complexity of a single iteration is $O(|V| + (k+ou+hl)*log|V| + k^3 + k^2 * DOF + k*R*min(\frac{1}{1-\alpha}, maxlen) + k*PR\_UPDATE)$. Since there are $\frac{B}{k}$ iterations, the total time complexity is $O(\frac{B}{k}*|V| + B*\frac{(k+ou+hl)}{k}log|V| + k^2 * B + k*B*DOF + B*R*min(\frac{1}{1-\alpha}, maxlen) + B*PR\_UPDATE)$.

\section{Experimental Evaluation}
All algorithms have been implemented in C++17 using GCC
7.5.0 compiler on a system running Ubuntu 18.04.5 with a 2.1 GHz
Intel(R) Xeon(R) Platinum 8160 processor and 256 GB RAM.

\subsection{Datasets}
For the real-world datasets, we use the Amazon product review dataset released by ~\cite{he2016ups}. It comprises customer reviews and ratings for different Amazon products. For the purpose of this paper we use dataset of four different products. We apply the rating-SVD algorithm as described in ~\cite{dash2022fairir} on these datasets to find the embedding and the top-10 RIRs for each item, and then create RIN. The details of RIN for each of the four datasets has been summarised in Table 1. We later use these embeddings to determine similarity score between items.

We observed that the generated RIN's are scale-free networks i.e node in-degrees follow a power law distribution. Figure 1 shows the plot of log of frequency vs log of in-degree. From the figure we can observe that Clothing RIN has relatively large number of hubs as compared to Electronics. We further observed that Books and Sports RIN have similar plot as Electronics.

\begin{table}
    \centering
\begin{tabular}{ |c|c|c|c|c|}
 \hline 
   \textbf{Name}  & $|V|$ & $|E|$ & $\ell_1$-norm & \%Over-Exp  \\
   \hline
    \textbf{Electronics} & 63001 & 630010 & 1.3669 & 8.2\\
     \hline
    \textbf{Books} & 367982 & 3679820 & 1.0834 & 18.1 \\
   \hline
    \textbf{Sports} & 957764 & 9577640 & 1.0475 & 22.4\\
   \hline
    \textbf{Clothing}  & 2681297 & 26812970 & 1.6991 & 0.05 \\
 \hline
\end{tabular}
    \caption{Dataset Description and Statistics }
    \label{tab:my_label}
\end{table}

\begin{figure}
    \centering
    \subfloat[\centering Amazon Electronics]{{\includegraphics[width=3.8cm]{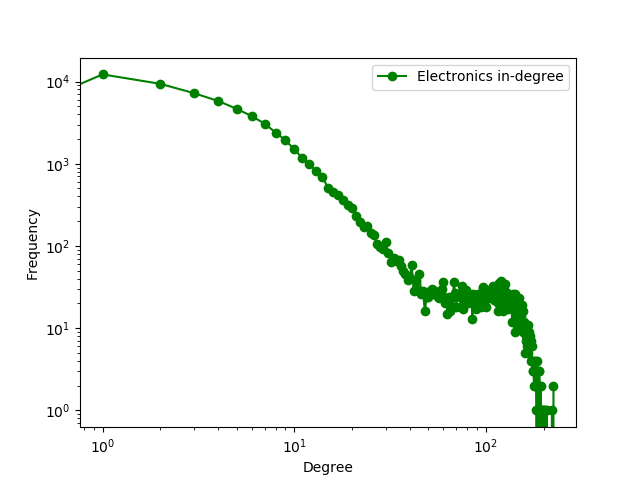} }}%
    \qquad
    \subfloat[\centering Amazon Clothing]{{\includegraphics[width=3.8cm]{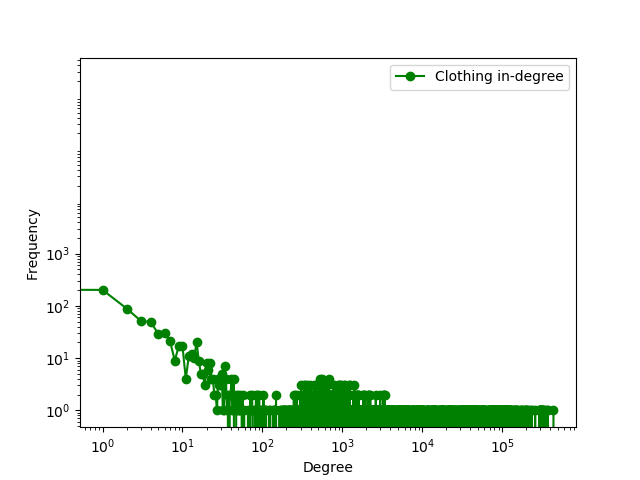} }}%
    \caption{Node In-degree Distribution}%
    \label{fig:example}%
\end{figure}

\subsection{Target Score Vector}
To compute the target score vector for a RIN we assume the average user-rating of an item as the quantification of its merit/quality and use the method in ~\cite{dash2022fairir} to estimate the desired exposure of an item. Let the average user-rating of $i^{th}$ item be $\boldsymbol{u}_i$ and its desired absolute PageRank score be $\boldsymbol{s}_i$. Then we compute $\boldsymbol{s}_i = C(\frac{\beta}{n} + (1-\beta)\frac{\boldsymbol{u}_i}{n})$ where $n$ is the number of nodes in RIN, C is the normalization over all vertices and $\beta \in [0,1]$ is a parameter that controls the extent of importance that we attach to the merit/quality of an item. For a chosen target set $T$, we use $\boldsymbol{t}=\boldsymbol{s}_T$ without renormalizing the selected scores on $T$. In Table 1, we have used $\beta=0$ to list the  \%over-exposed nodes and the initial $\ell_1$-norm between the original PageRank vector and the target score vector of the respective RINs. Observe that relative to other RINs, Clothing RIN has very less fraction of over-exposed nodes though their contribution to $\ell_1$-norm is significant i.e they have much higher PageRank than desired.

\subsection{Performance metric and parameters}

\subsubsection{Performance Metrics}
To quantify the performance of our heuristics, we use the following metrics:
\begin{itemize}
    \item \bm{$\ell_1$} \textbf{norm}: For the addition multiset $E'$ returned by a heuristic, we measure $\ell_1(\boldsymbol{\tilde{\pi}}_T(E'),\boldsymbol{t})$ on the target vertices.
    \item \textbf{Average Similarity}: To measure the relevance of an edge $(u, v)$ added by the heuristic, we compute the similarity score for it as the normalised dot product of the item embeddings represented by nodes $u$ and $v$. Then we take the average of the similarity score of all edges added by the heuristic. 
\end{itemize}

\subsubsection{Input Parameters}
The $\ell_1$ \textsc{Target PageRank} problem consists of the following input parameters:
\begin{itemize}
    \item \textbf{Budget}: We measure the performance of the heuristics for the budget values in the set $\{0.5\%,\ 1\%$, $1.5\%\}$ of $|E|$.
    \item \textbf{Target Nodes}: We use two different target set of nodes:
    \begin{enumerate}
        \item \text{2-HOP}: First randomly select 10\% of the nodes from $V$ and add all nodes at a distance of atmost 2 from them in $G$. Adding 2-hop neighbours helps in representing related items. 
        \item \text{WHOLE}: Take $V$ as the target set.
    \end{enumerate}
    \item \textbf{Target Score Vector}: To generate different target score vectors, we use the following values of $\beta$: $\{0.0,\ 0.5,\ 1.0\}$.

    \item \textbf{Default Parameters}: To compute the PageRank values and perform random walks we set $\alpha=17/20$, giving teleportation probability $1-\alpha=3/20=0.15$ for all experiments. Unless mentioned, we have used Amazon Books dataset with budget as 1\% of $|E|$, $\beta = 0$ for the target score vector and target set as the whole vertex set.
\end{itemize}

\subsubsection{Hyper Parameters}
The heuristics involve the following hyperparameters:
\begin{itemize}
    \item \textbf{k}: This parameter controls the number of edges added in a single iteration. We have used the following values of $k$ for our experiments: $\{0.066\%,\ 0.1\%,\ 0.2\%\}$ of $|V|$.
    \item \textbf{ou}: This parameter controls the number of over-exposed and under-exposed nodes used by the heuristics. For G2 and RWH, we use the following values of $ou$: $\{0.1\%,\ 1\%,\ 2\%,\ 4\%,\ 8\%\}$ of $|V|$. Whereas for G1 we use the following values of $ou$: $\{60\%,\ 75\%,\ 90\%,\ 95\%, \newline \ 100\%\}$ of $k$.
    \item \textbf{hl}: This parameter controls the number of high PageRank and low PageRank nodes used by the heuristics. For G2 and RWH, we use the following values of hl: $\{1,\ 2,\ 4,\ 8, \ 16, \ 32, \ 48 \}$ $\times$ $k$. Whereas for G1, we have $hl = k - ou$.
\end{itemize}
We run a grid search on the above parameters by varying them as specified and obtain the set of parameters that give the best results.

\subsection{Comparison of Heuristics}
\subsubsection{Variation With Input Parameters}
We study the performance of heuristics on varying input parameters.
\begin{itemize}
    \item \textbf{Budget}: We evaluate the performance of heuristics against varying the budget. Figure 2 displays the results for Amazon Books dataset for the \text{2-HOP} target set with budget values mentioned earlier. We notice that on increasing the budget, all 3 heuristics decrease the $\ell_1$-norm. However, we notice a saturation in the decrease, i.e. we see a decrease in the rate of change of $\ell_1$-norm on increasing the budget. We also observe that the saturation occurs faster in G1 and G2 heuristics as compared to RWH heuristic. We observe similar results for all datasets.
    
    \begin{figure}%
        \centering
        \subfloat[\centering]{{\includegraphics[width=6cm]{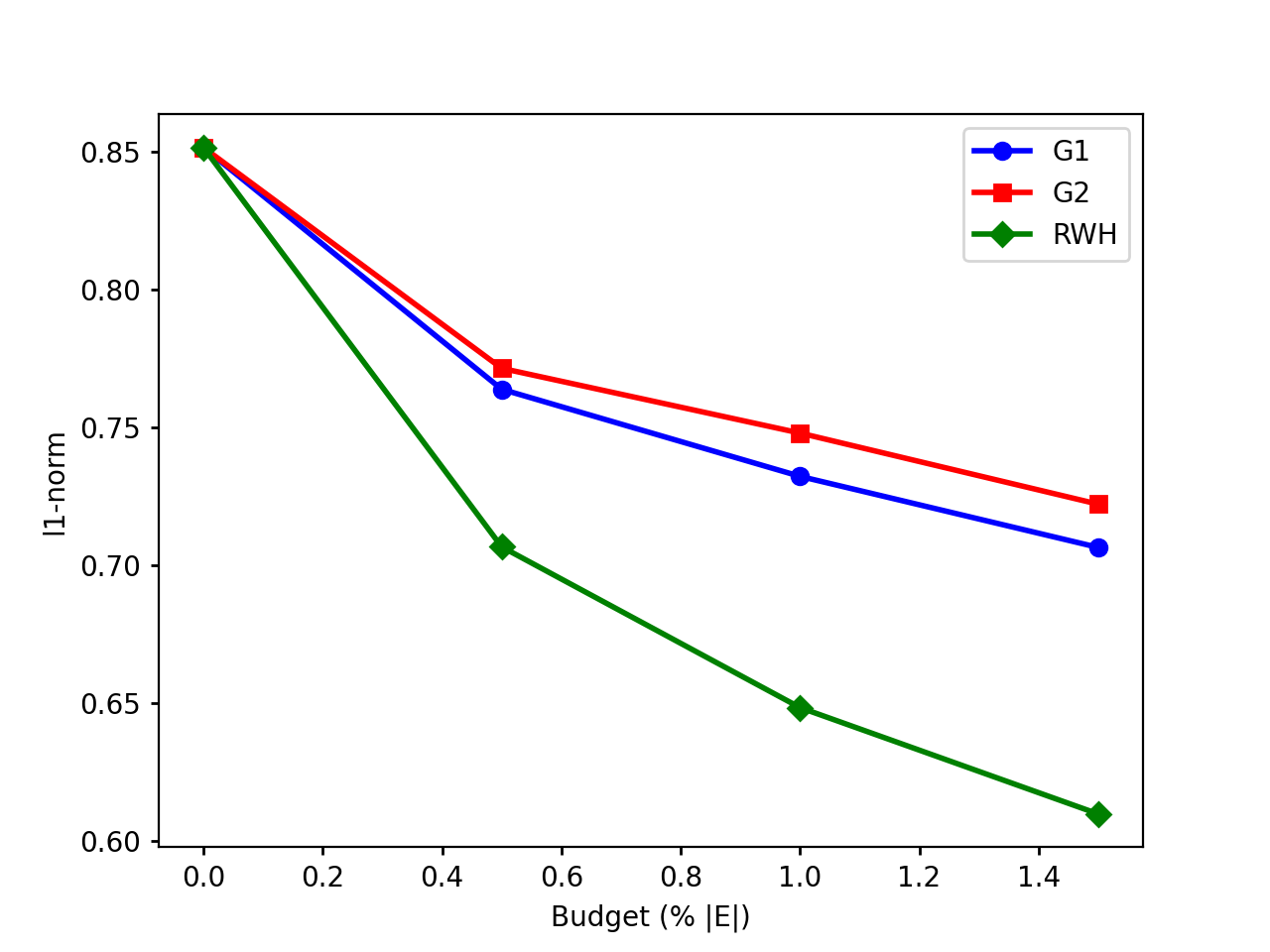} }}%
        \caption{Variation With Budget}%
        \label{fig:example}%
    \end{figure}
    
    \item \textbf{Target Nodes}: We analyze the effect of changing target nodes on the performance of heuristics using the \text{2-HOP} and \text{WHOLE} target sets on all datasets. The results are displayed in Figure 3 where in the x-axis, I is used to represent the initial $\ell_1$-norm and G1, G2 and RWH are used to represent the final $\ell_1$-norm after running the respective heuristics. We observe that in every dataset except Clothing, RWH performs better than G1 and G2, whereas the performance of G1 is slightly better than G2. This trend is violated only in Clothing dataset when the target set is \text{2-HOP}. This is explainable as in the Clothing dataset there are a very small number of over-exposed nodes and they have extremely high in-degree and PageRank. So, it is highly likely that these nodes, along with the ones they are connected to are contained in the \text{2-HOP} target set. So, \text{2-HOP} essentially filters out the clusters in the graph and the over-exposed nodes become the cluster centers. So, RWH has no advantage over other 2 heuristics in terms of computing cluster centers to add edges. Secondly, we observed that the over-exposed nodes have so high PageRank that adding edges from them to under-exposed nodes causes the PageRank of under-exposed nodes to shoot up much beyond their target, thereby increasing the $\ell_1-$norm. In this scenario, G2 heuristic performs better as it adds edges greedily to reduce $\ell_1$-norm while RWH adds edges based on node similarity. Hence, G2 is expected to perform better.
\begin{figure}%
    \centering
    \subfloat[\centering Target: \text{2-HOP}]{{\includegraphics[width=6cm]{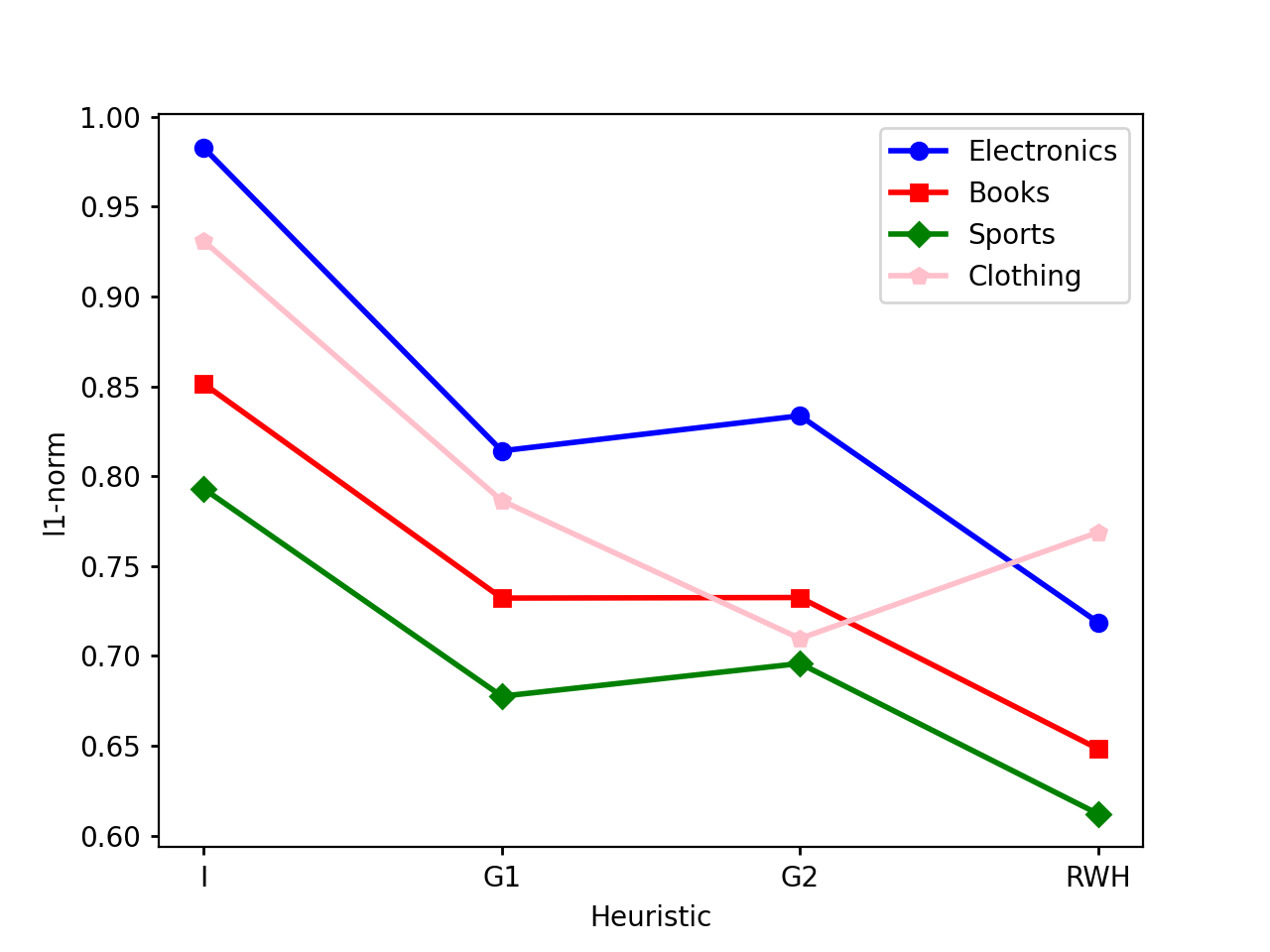} }}%
    \qquad
    \subfloat[\centering Target: \text{WHOLE}]{{\includegraphics[width=6cm]{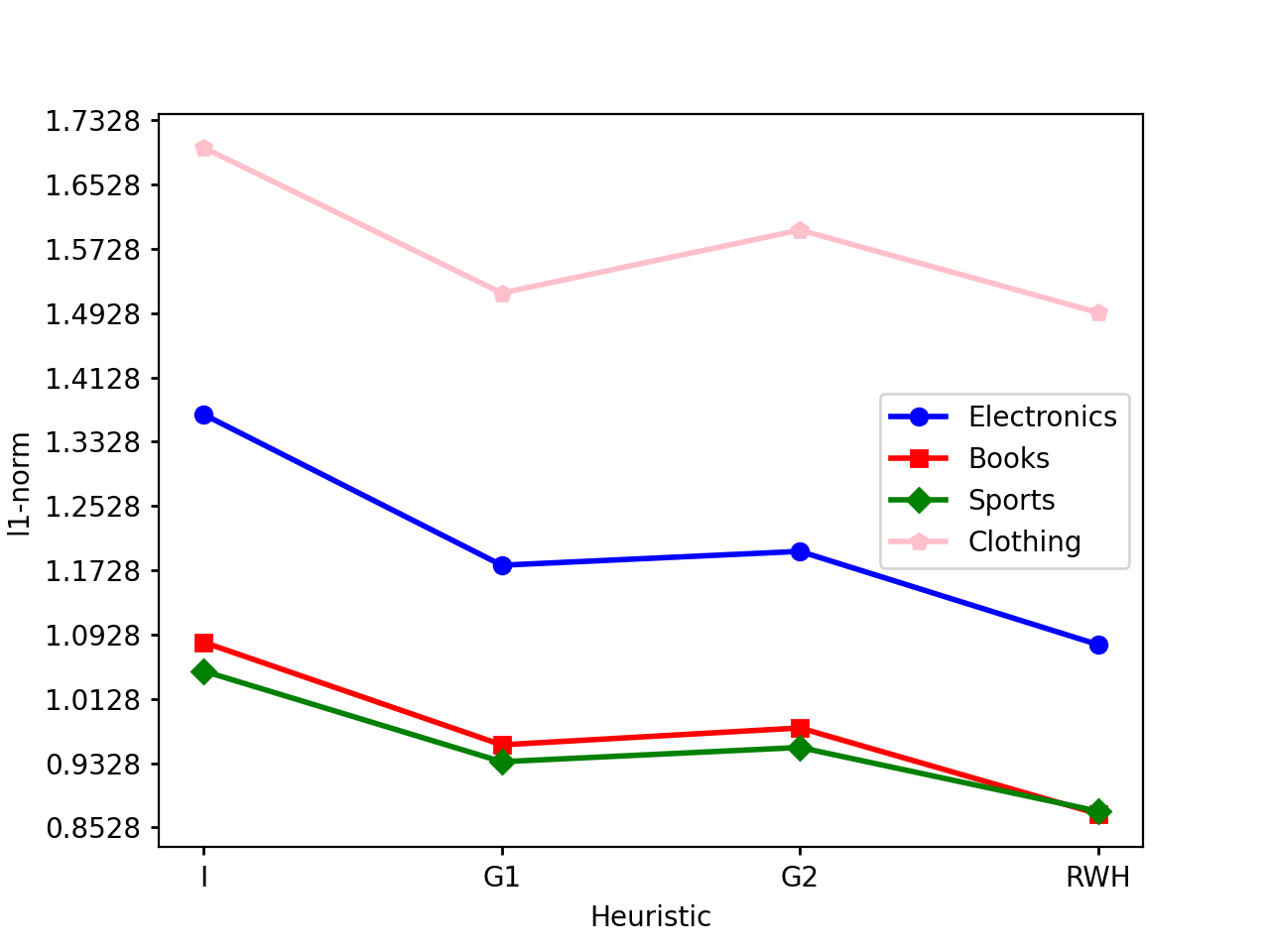} }}%
    \caption{Variation with Target Sets}%
    \label{fig:example}%
\end{figure}    
    \item \textbf{Target Score Vector}: Figure 4 shows the effect of changing $\beta$ on the heuristics. The figure shows the plot of logarithm of $\ell_1$-norm with $\beta$. We observe that on increasing $\beta$, the initial $\ell_1$-norm first decreases, then increases. From this, we can conclude that there a significant number of over-exposed (under-exposed) nodes with PageRank less (greater) than $\frac{1}{n}$ in our dataset. We also observe that all heuristics decrease the logarithm of $\ell_1$-norm by similar amount in each case implying the percentage change in $\ell$-norm is similar.
\begin{figure}%
    \centering
    \subfloat[\centering]{{\includegraphics[width=6cm]{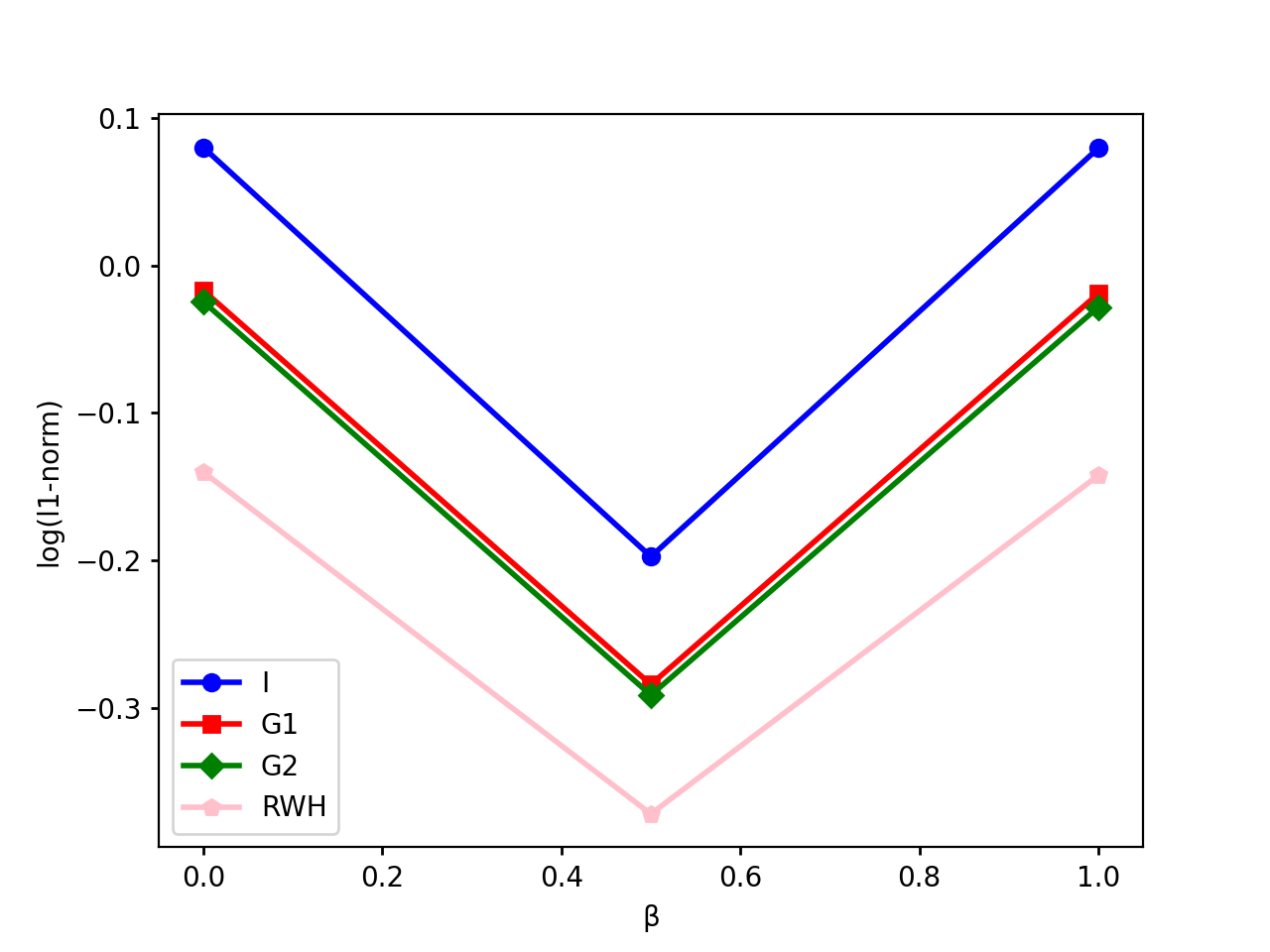} }}%
    \caption{Variation with $\beta$}%
    \label{fig:example}%
\end{figure}   
\end{itemize}

\subsubsection{Effect on Average Similarity}
 We analyse the relevance of edges added by different heuristics. Figure 5 shows the plot of Average Similarity against different heuristics for Electronics and Books dataset. We observe that RWH performs much better than the other 2 heuristics. This is because RWH finds a maximum weight matching based on edge similarity score as the weight. On the other hand, the other 2 heuristics don't add edges based on edge similarity.
 
\begin{figure}%
    \centering
    \subfloat[\centering]{{\includegraphics[width=6cm]{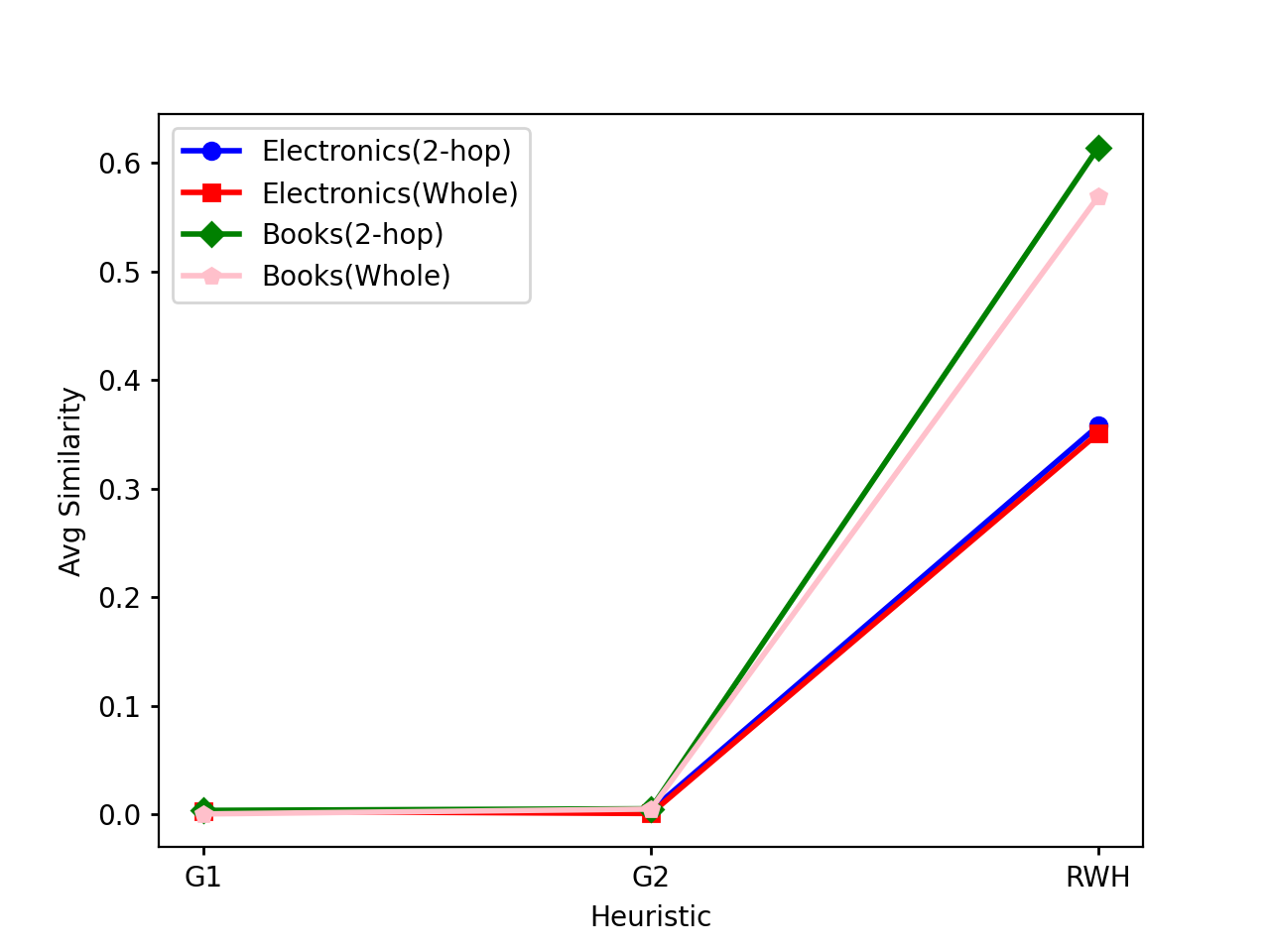} }}%
    \caption{Average Similarity}%
    \label{fig:example}%
\end{figure}

\subsubsection{Comparision on Scalability}
Figure 6 shows the plot of logarithm of time taken by each heuristic to run v/s the logarithm of the number of nodes in each RIN. We observe that time taken by all heuristics is almost linear with the graph size. Hence, they all scale well. However, we observe that G2 takes significantly more time than RWH, and RWH takes slightly more time than G1. This is expected because in G2, to choose the edges greedily we need to compute the change in $\ell_1$-norm for each of them, whereas in RWH no such computation is required and finding cluster centers and max matching is much cheaper. On the other hand, G1 only requires sorting nodes based on their over/under-exposure and their PageRank values before adding edges, making it least computationally expensive.

\begin{figure}%
    \centering
    \subfloat[\centering]{{\includegraphics[width=6cm]{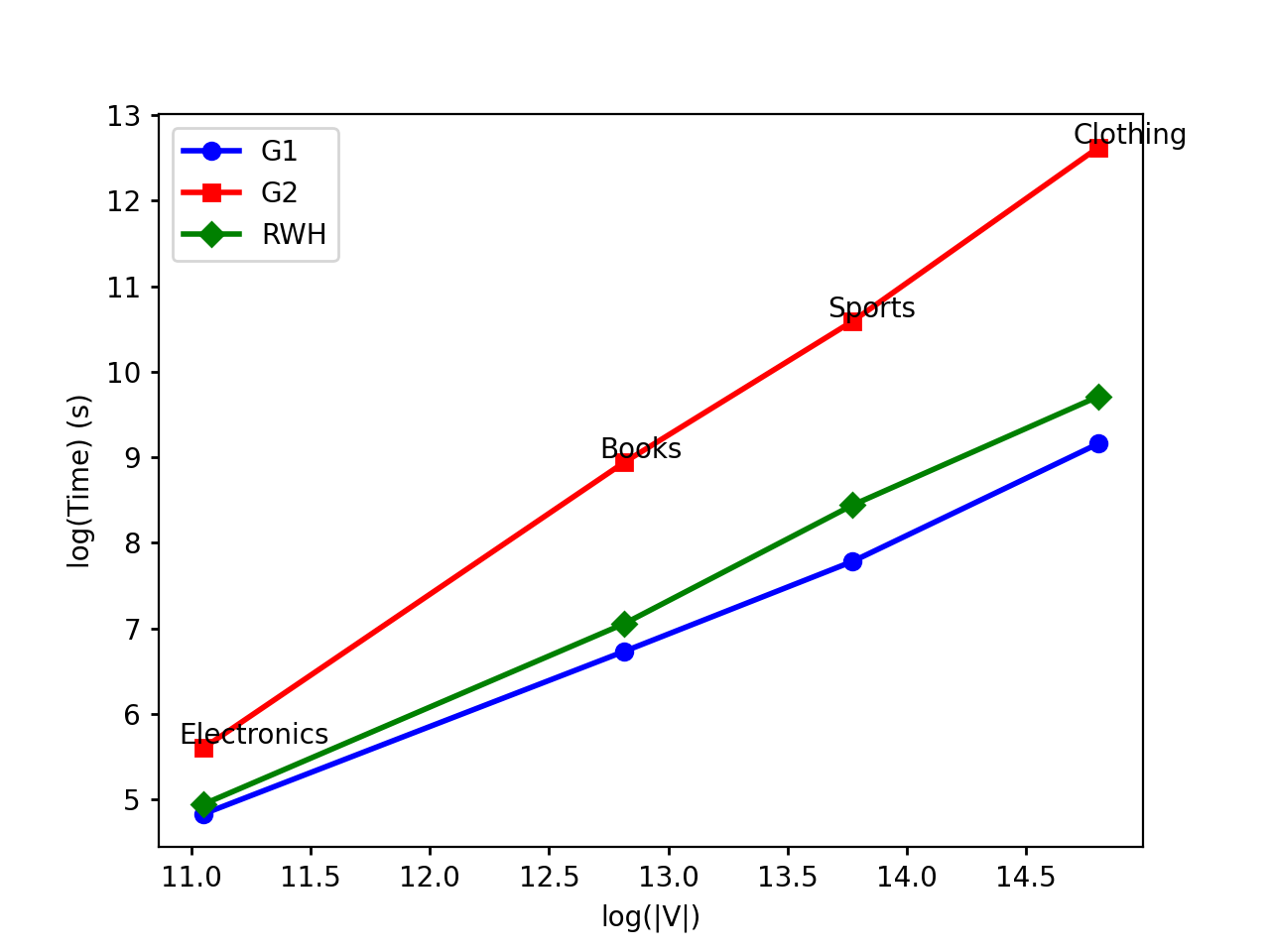} }}%
    \caption{Time Taken}%
    \label{fig:example}%
\end{figure}
\subsection{Effect of Hyper-Parameters}
We study the effect of hyper-parameters on the performance of heuristics.

\subsubsection{Variation with k}
In Figure 7(a) we observe that on increasing $k$, the average similarity for RWH heuristic increases. This is because more nodes are present for matching and there is more scope to increase total similarity by finding maximum weight matching. We didn't observe any noticeable change for G1 and G2 since these heuristics don't consider edge similarity while adding edges. \\
In Figure 7(b) we observe that on increasing $k$, the time taken by the RWH decreases. This is because finding maximum matching plays a significant role in the total time taken by RWH heuristic. There was no significant impact of k on the time taken by $G1$ and $G2$, because the the computationally expensive steps in them are independent of $k$. \\
On changing $k$ within the specified range, there was no noticeable trend in change of $\ell_1$-norm in any of the heuristics.

\begin{figure}%
    \centering
    \subfloat[\centering Average Similarity]{{\includegraphics[width=6cm]{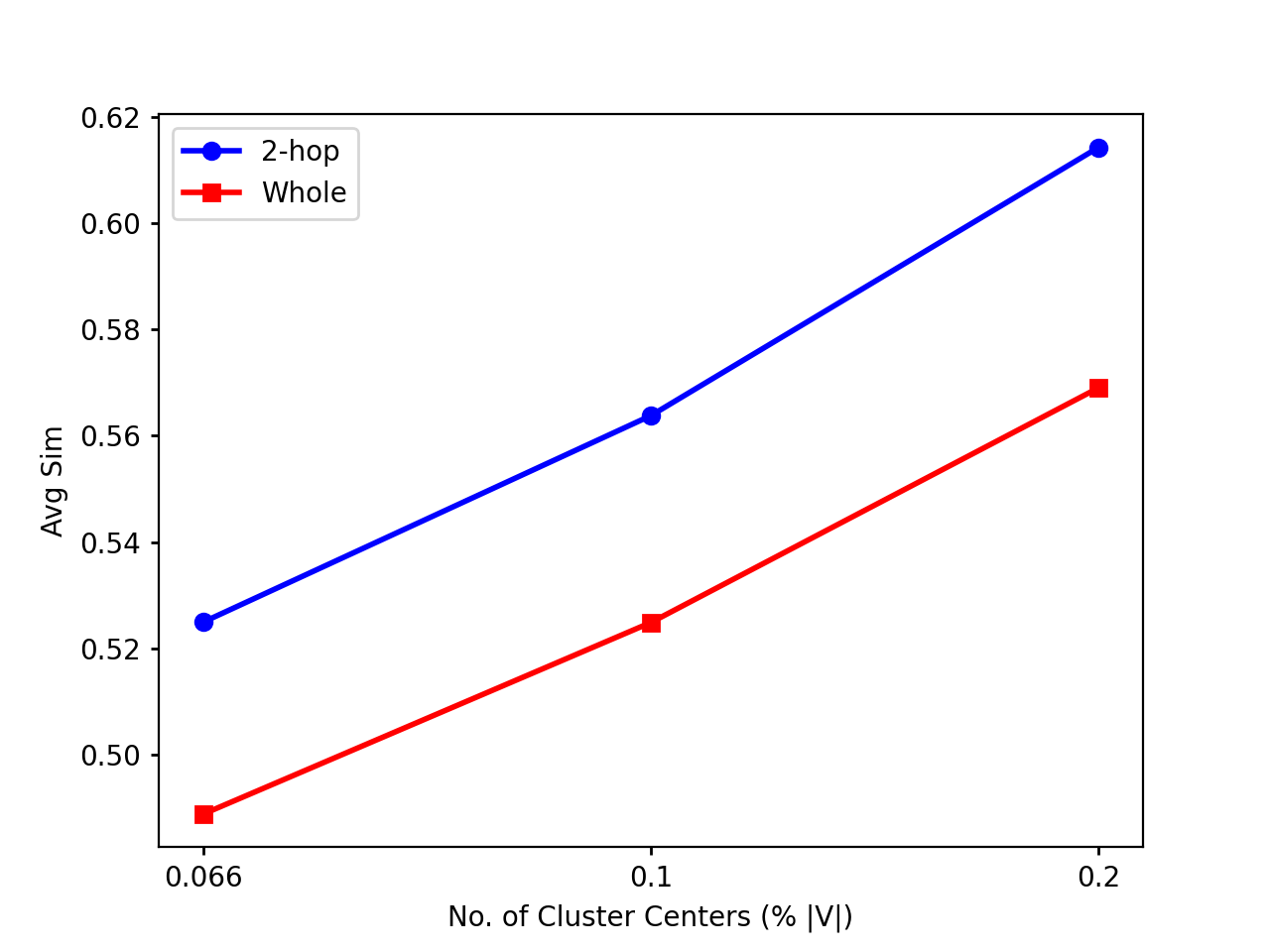} }}%
    \qquad
    \subfloat[\centering Time Taken]{{\includegraphics[width=6cm]{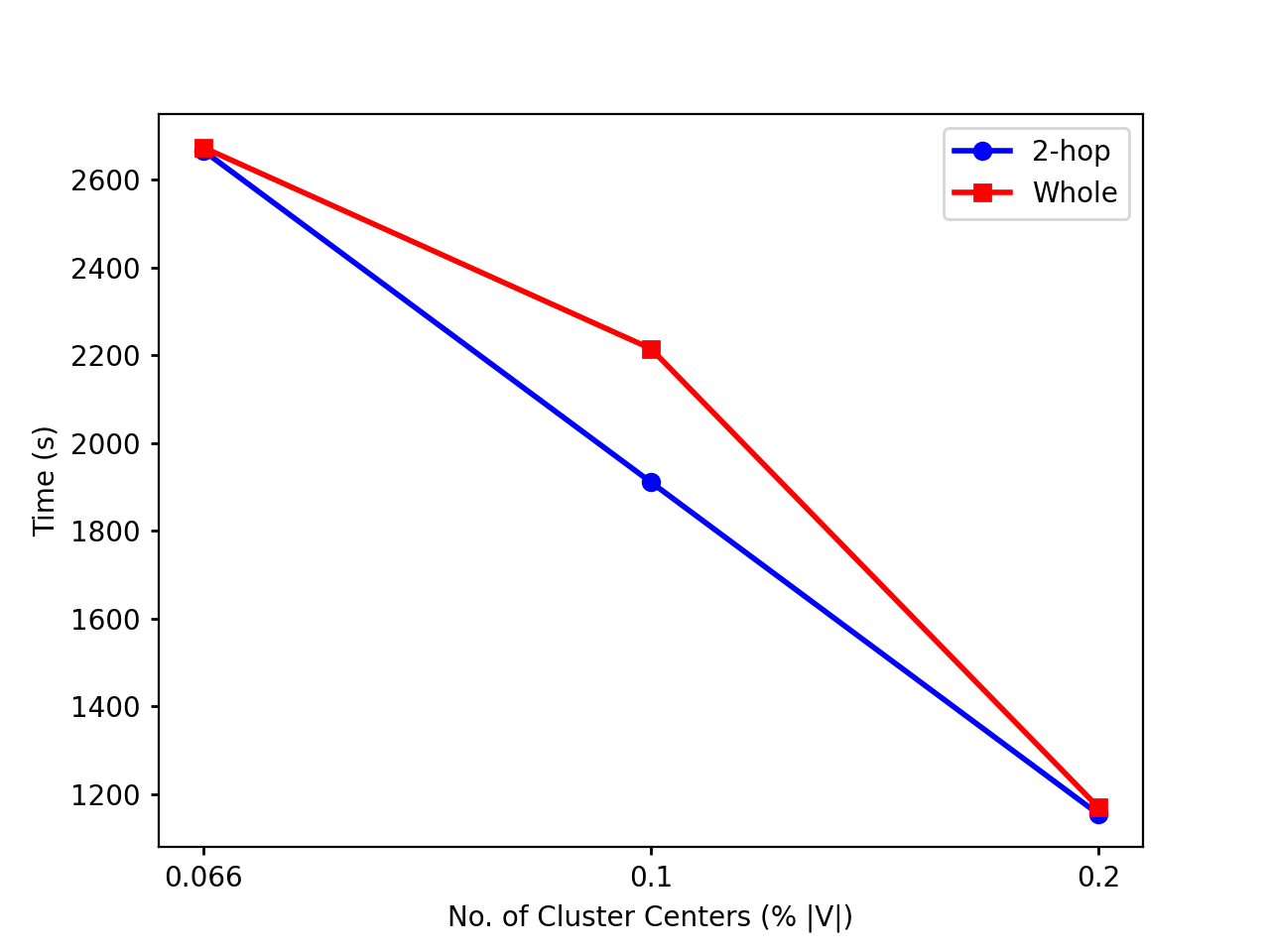} }}%
    \caption{Variation with $k$ (No. of Cluster Centers)}%
    \label{fig:example}%
\end{figure}

\subsubsection{Variation With ou}
Figure 8 displays the performance of G1, G2 and RWH heuristic for various values of $ou$ on Amazon Books dataset. We observe in each case there is a sweet spot in $ou$, i.e. the $\ell_1$-norm first decreases, then increases. This can be explained as follows:
\begin{itemize}
    \item RWH finds clusters of over and under exposed nodes. When the value of $ou$ is too small, the heuristic doesn't get sufficient nodes to find good clusters. On the other hand, when the value of $ou$ is too high, many over/under exposed nodes are given to the heuristic and cluster finding is influenced by nodes that aren't too over/under exposed. This reduces the quality of cluster centers and hence decreases the performance of the heuristic.
    \item Since G1 and G2 choose edges from over-exposed to under-exposed nodes greedily, using too few $ou$ nodes will cause some good nodes to not be used for edge additions. As in the case of RWH, a high value of $ou$ will cause too many nodes to be picked because of which the quality of the random edges selected will suffer.

\end{itemize}
 
\begin{figure}%
    \centering
    \subfloat[\centering G1]{{\includegraphics[width=6cm]{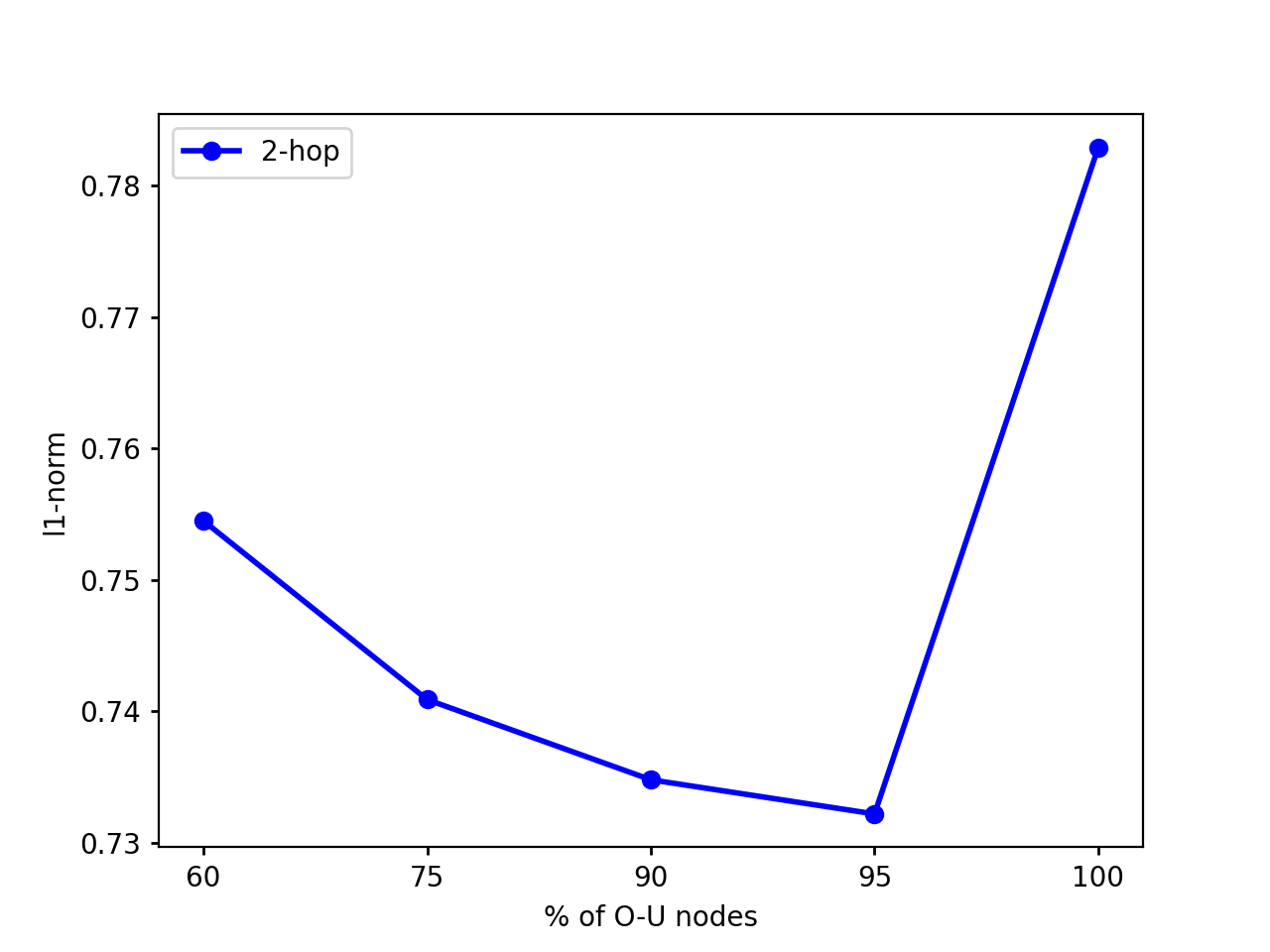} }}%
    \qquad
    \subfloat[\centering G2 and RWH]{{\includegraphics[width=6cm]{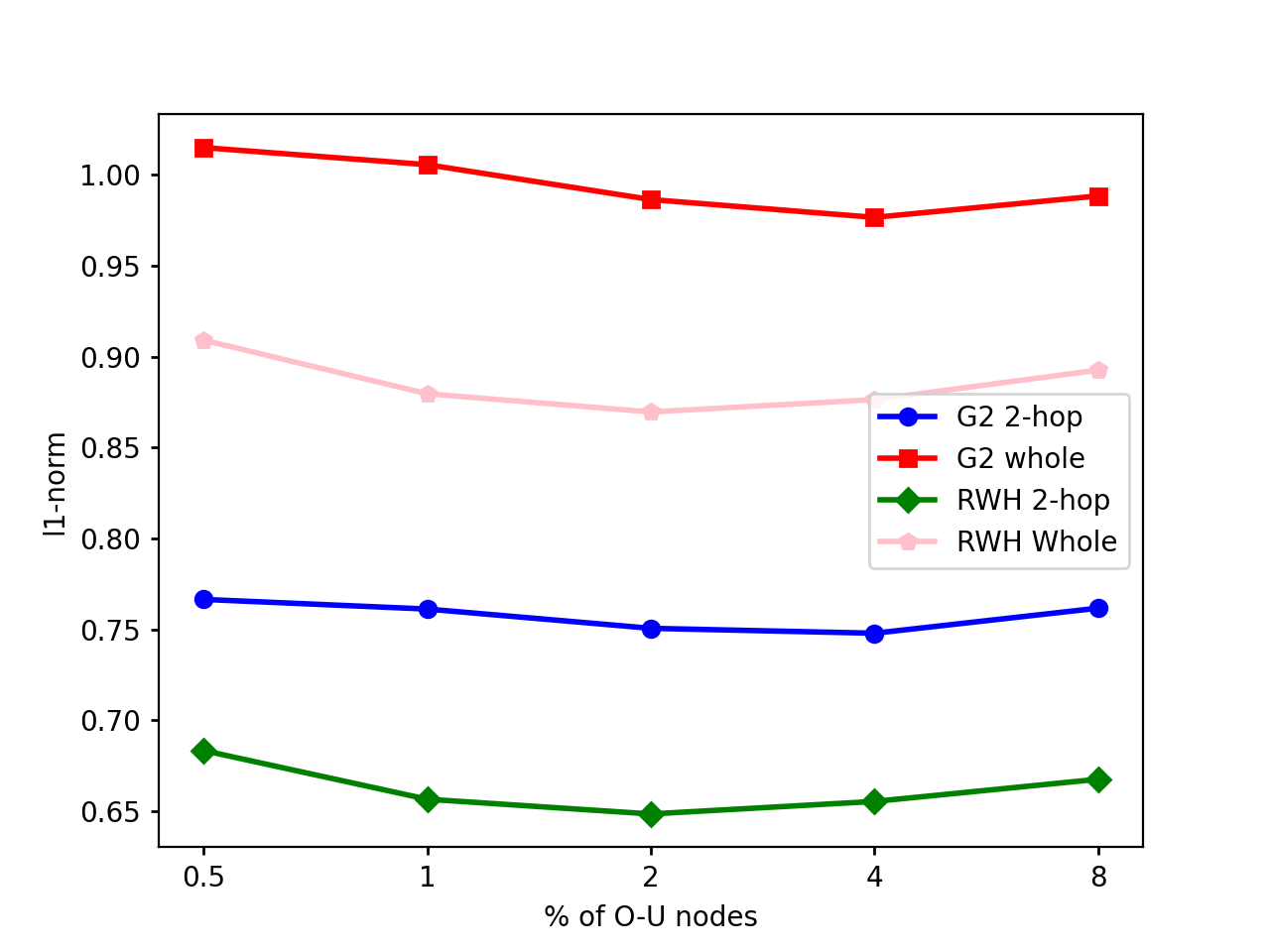} }}%
    \caption{Variation With ou}%
    \label{fig:example}%
\end{figure}

\subsubsection{Variation With hl}
Figure 9 displays the performance of G2 and RWH heuristic for various values of $hl$ on Amazon Books and Amazon Sports dataset on \text{2-HOP} target set. We observe that on changing $hl$, there is little effect on performance RWH. This signifies that the high/low PageRank nodes aren't as good cluster centers as over/under exposed nodes. This is explainable because in the \text{2-HOP} target set, an over-exposed (under-exposed) node is likely to be surrounded by other over-exposed (under-exposed) nodes increasing its chances to become a cluster center. Hence, the participation of $hl$ nodes in edge additions is less in RWH. We also see that the effect on G2 is more prominent and the trend is dataset dependent. This is because in some datasets like Clothing adding edges from nodes in target set (over/under-exposed nodes) to  nodes out of target set (high/low PageRank nodes) will have more reduction in $\ell_1$-norm than adding edges from over-exposed to under-exposed nodes. Hence the participation of $hl$ nodes in additions is more in G2.\\
Overall, we found that the effect of over/under-exposed nodes is more dominant than high/low PageRank nodes in each of the heuristics.

\begin{figure}%
    \centering
    \subfloat[\centering]{{\includegraphics[width=7cm]{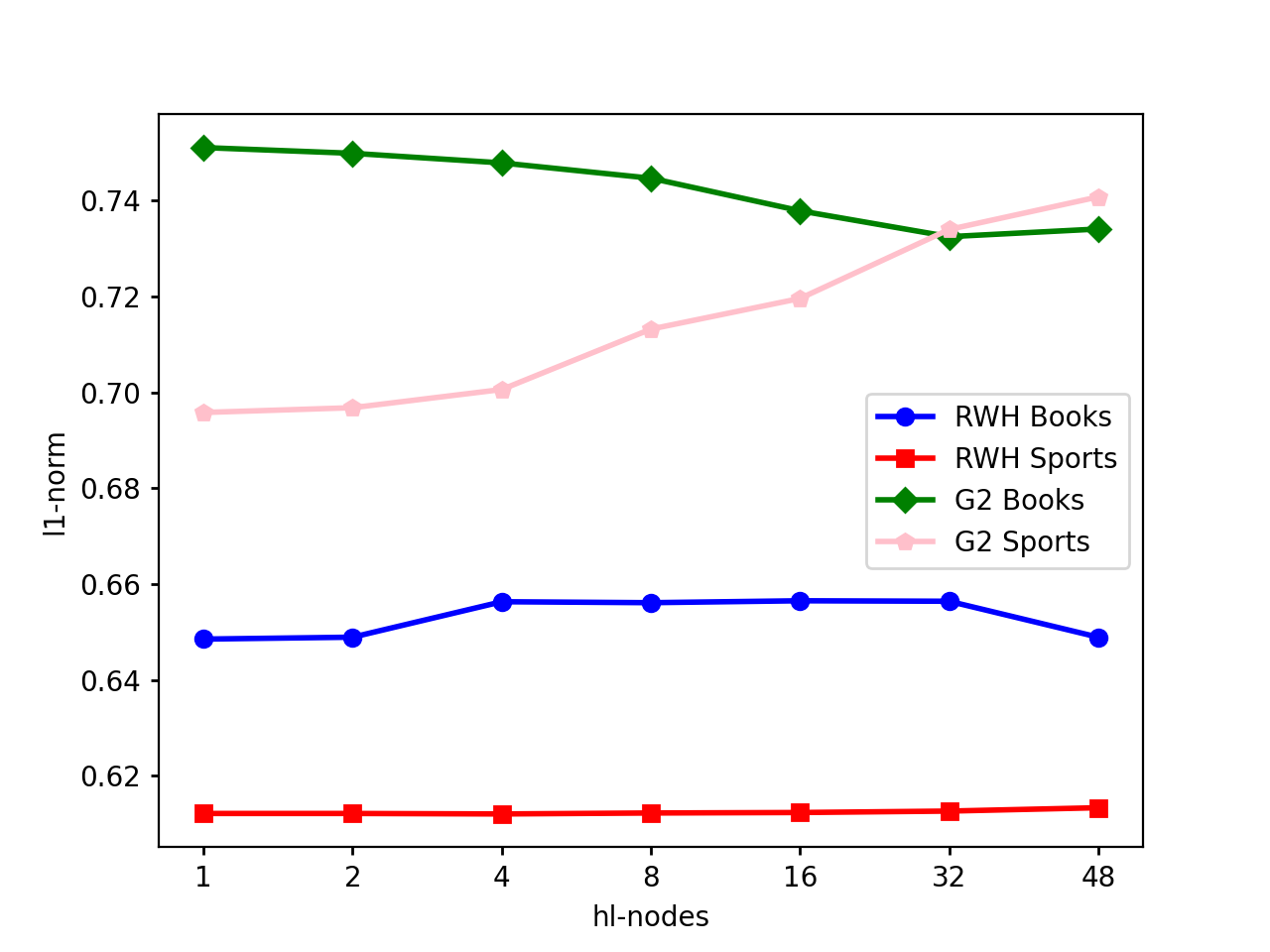} }}%
    \caption{Variation With hl}%
    \label{fig:example}%
\end{figure}

\subsection{Experiments on Synthetic Datasets}
We perform experiments on synthetic datasets to verify the consistency in performance of our heuristics.
\subsubsection{Graph Generation}
We have used the Barabasi-Albert model ~\cite{barabasi1999emergence} to generate directed graphs that follow power law for node in-degrees and have constant out-degree. We fix the number of nodes in the graph to be 50000 and the out-degree of each node as 10. The graph is generated as follows:
\begin{enumerate}
    \item Start with a network of 10 nodes (no edges).
    \item Add new nodes. For each new node added, add edges as follows:
    \begin{itemize}
        \item With probability $p \in [0, 1]$, add an edge from the new node to 10 randomly chosen existing nodes.
        \item With probability $1-p$, the new node connects to $m=10$ existing nodes chosen with probability proportional to their indegrees.
    \end{itemize}
\end{enumerate}
It is shown in ~\cite{kumar2000stochastic} that the above process generates a graph following power law with in-degree distribution $d(k) \sim k^{-\beta}$ where $\beta = \frac{2-p}{1-p}$, which implies that on increasing $p$ power law exponent increases thereby making the graph more uniform.

\subsubsection{Target Score Vector Generation}
We generate scores on all vertices by adding Gaussian noise to their original PageRank as follows:
\begin{enumerate}
    \item Compute the original PageRank distribution $\boldsymbol{\pi}$ of the graph $G(V, E)$.
    \item Sample i.i.d random variables $X_v \sim \mathcal{N}(0, 1)$ for each node $v \in V$.
    \item Compute the score vector $\boldsymbol{s}$ where $\boldsymbol{s}_v = C(\boldsymbol{\pi}_v + \lambda |X_v|)$, with $C$ normalizing over all of $V$ and $\lambda$ controls the weight of noise added.
\end{enumerate}

For either target set, use $\boldsymbol{t}=\boldsymbol{s}_T$ without renormalizing on $T$. We now study the effect of varying the power law exponent and target score vector:
\begin{itemize}
    \item \textbf{Varying} $\boldsymbol{p}$: We use $p=0.1$ and $p=0.5$ keeping $\lambda = 8 \times 10^{-5}$. The results are shown in Figure 10. We observe that in \text{2-HOP} target, RWH performs worse than G2 for $p=0.1$, whereas performance of G1 is mostly similar to G2. Notice that on decreasing the value of $p$, the graph will display a sharper power law exponent. Hence, there will be a small number of nodes with high in-degree and the graph becomes similar to Clothing dataset. So, the reason for the observed behaviour is similar to the one described for Amazon Clothing in Section 8.4. \\
    We also observe that for $p=0.1$, both G1 and G2 perform better than RWH. This further verifies the fact that RWH doesn't get an advantage in finding clusters using random walks due to the structure of the graph. Choosing edges based on $\ell_1$-norm gives G1 and G2 an advantage over RWH in this case.

    \item \textbf{Varying} $\boldsymbol{\lambda}$: We use $\lambda$ in the set $\{2 \times 10^{-5},\ 4 \times 10^{-5},\ 8 \times 10^{-5}\}$ keeping $p = 0.5$. Figure 11 shows the plot of logarithm of $\ell_1$-norm with the heuristic used. On increasing $\lambda$, we observe that  the initial $\ell_1$-norm increases as we are increasing the weight of noise in computing the target score vector. However, for each heuristic, the decrease in logarithm of $\ell_1$-norm is almost the same for each $\lambda$ implying the percentage decrease in $\ell_1$-norm to be nearly same.

\end{itemize}

\vspace{5mm}
\begin{figure}%
    \centering
    \subfloat[\centering Target: \text{2-HOP}]{{\includegraphics[width=3.8cm]{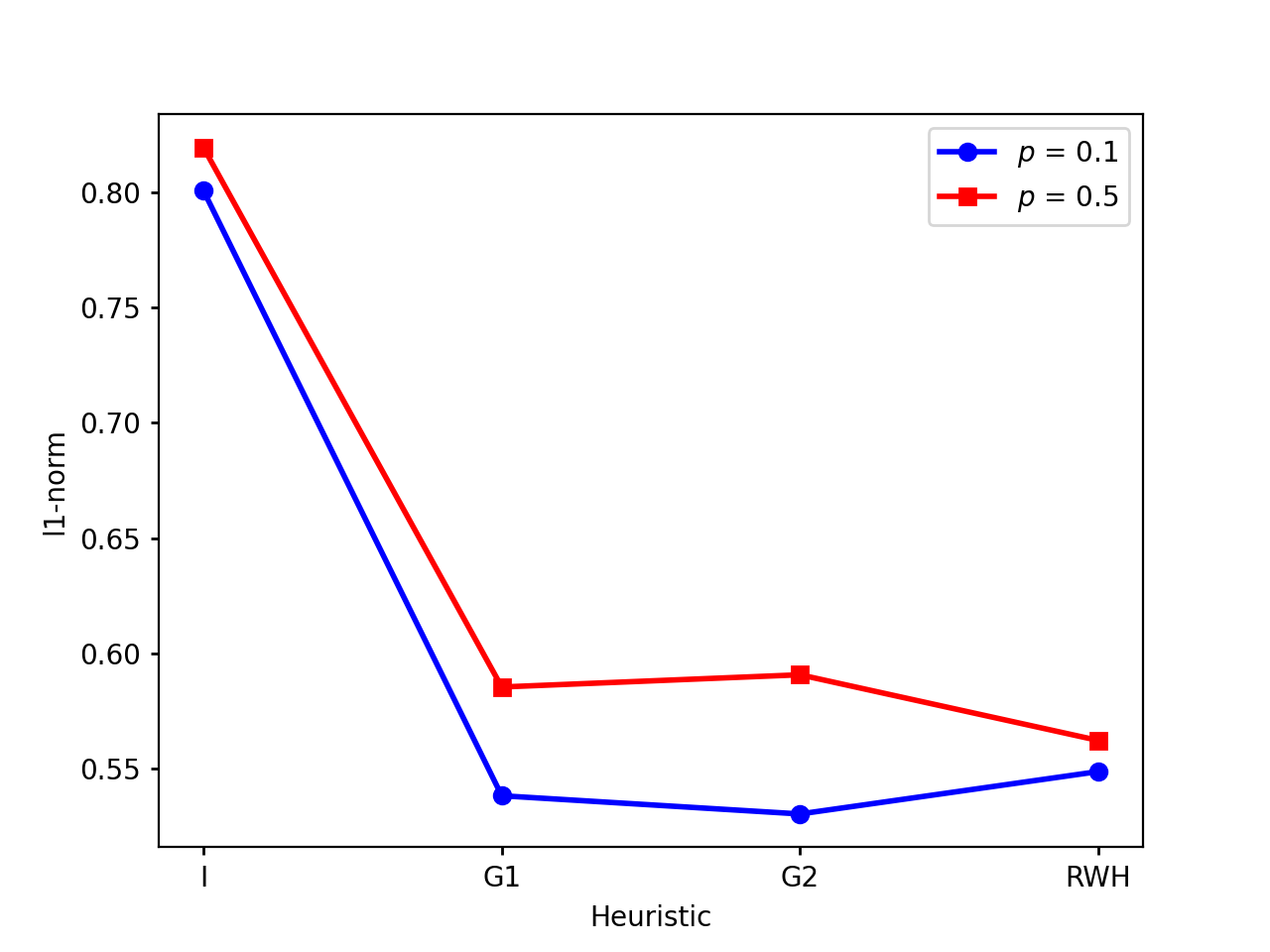} }}%
    \qquad
    \subfloat[\centering Target: \text{WHOLE}]{{\includegraphics[width=3.8cm]{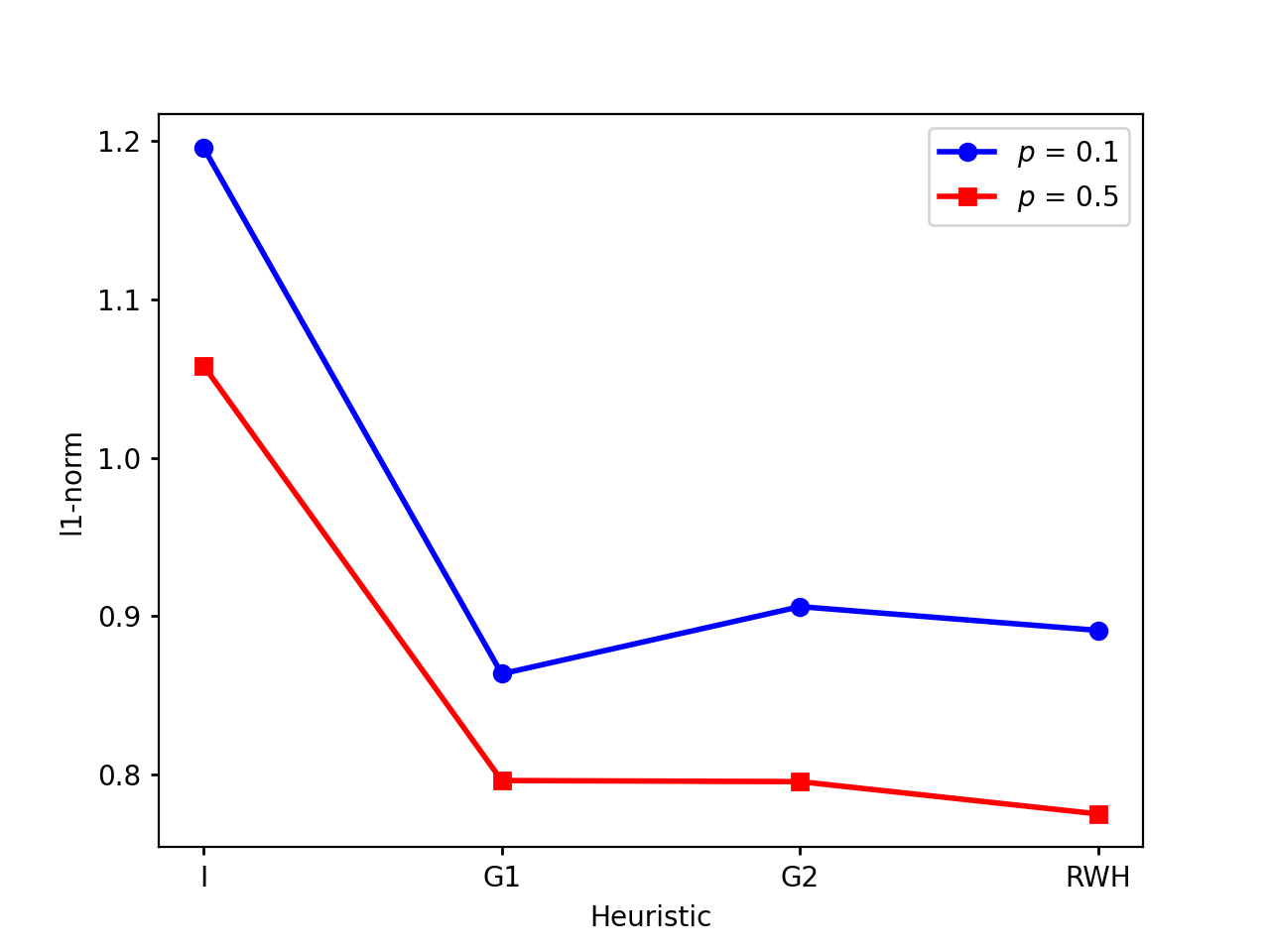} }}%
    \caption{Variation with $p$}%
    \label{fig:example}%
\end{figure}
    
\begin{figure}%
    \centering
    \subfloat[\centering Target: \text{2-HOP}]{{\includegraphics[width=3.8cm]{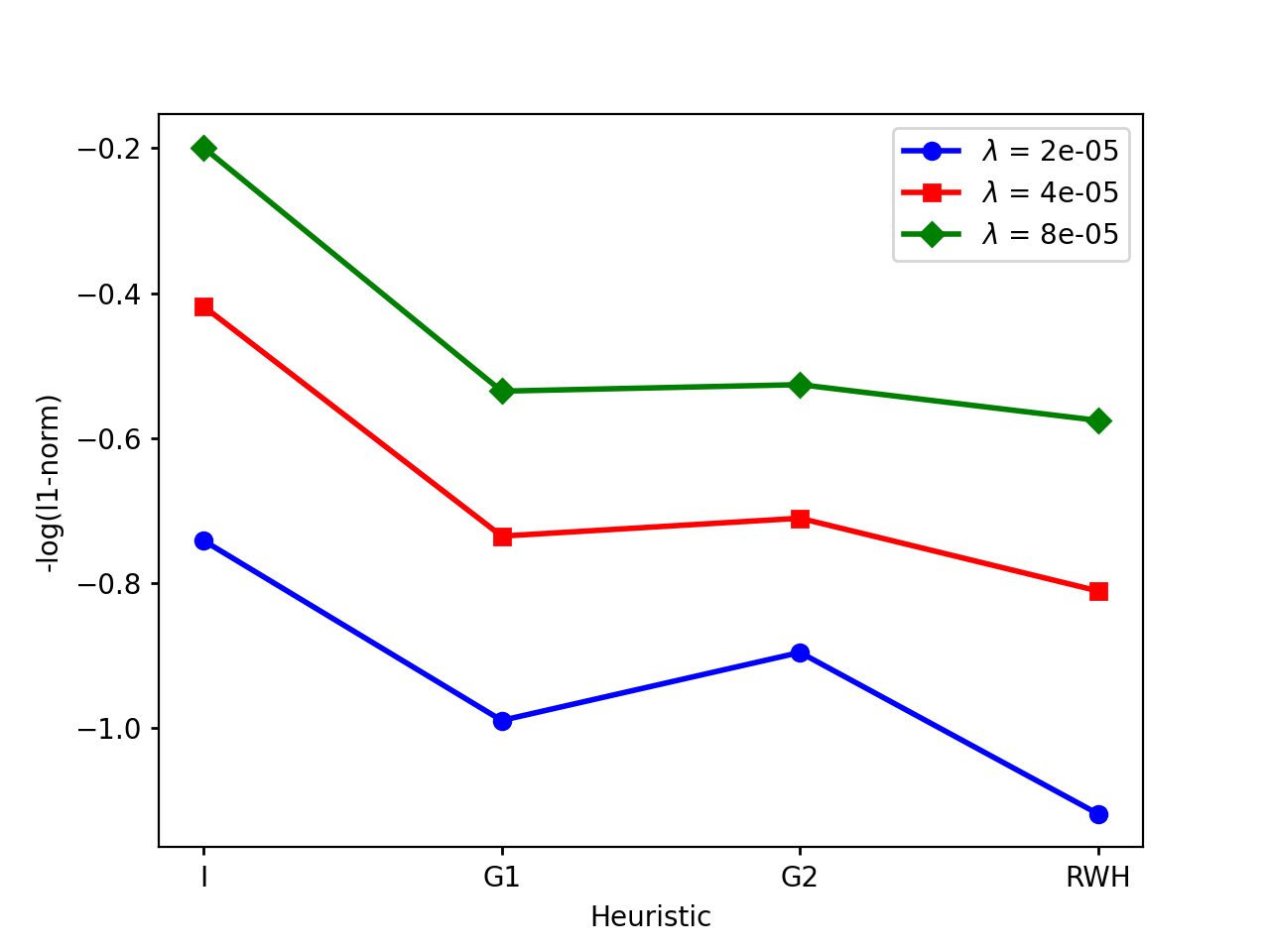} }}%
    \qquad
    \subfloat[\centering Target: \text{WHOLE}]{{\includegraphics[width=3.8cm]{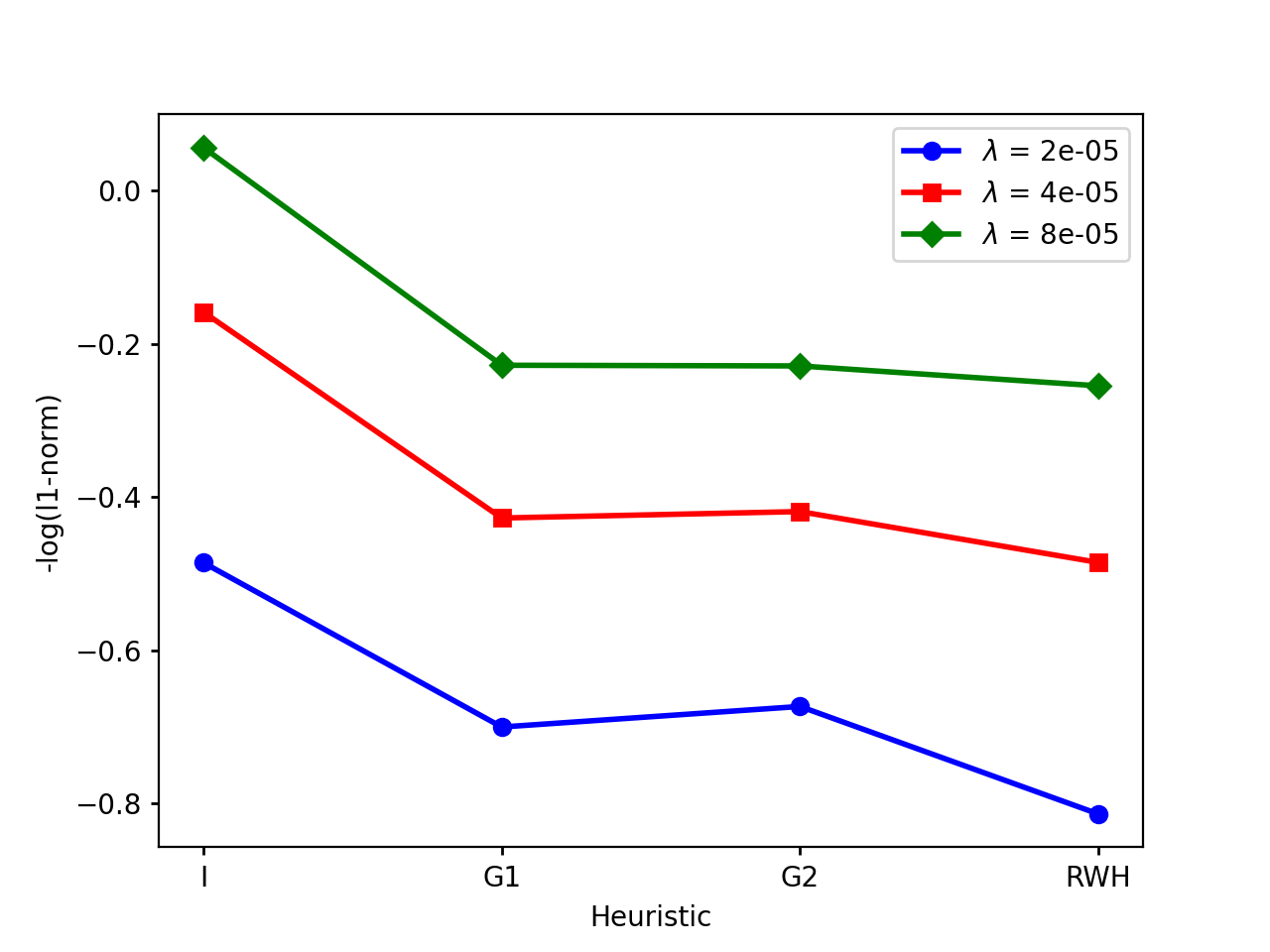} }}%
    \caption{Variation with $\lambda$}%
    \label{fig:example}%
\end{figure}
\section{Conclusion}
In this paper, we studied the problem of modifying the PageRank values of a fixed set of target nodes to match a given target PageRank vector, subject to a budget of exactly $B$ edge additions. We established that the corresponding decision problem is NP-complete. We then showed that minimizing the $\ell_1$-distance between the resulting and desired absolute PageRank scores on the target nodes admits no polynomial-time multiplicative constant-factor approximation unless $P=NP$. We also formulated the problem as a mixed-integer linear program (MILP), while noting the scalability limitations of this approach. Motivated by these computational challenges, we proposed multiple heuristics, including methods that incorporate node similarity when selecting edges to add, and evaluated their performance on multiple Amazon datasets.

In future work, we aim to compare the performance of our heuristics against the optimal MILP solution on smaller graphs. We also plan to apply our heuristics to item recommendation and investigate whether the modified recommendations better reflect user preferences, as measured through user reviews, when the target set of nodes and target PageRank vector are chosen appropriately.

\section{Proof of Theorem 1}
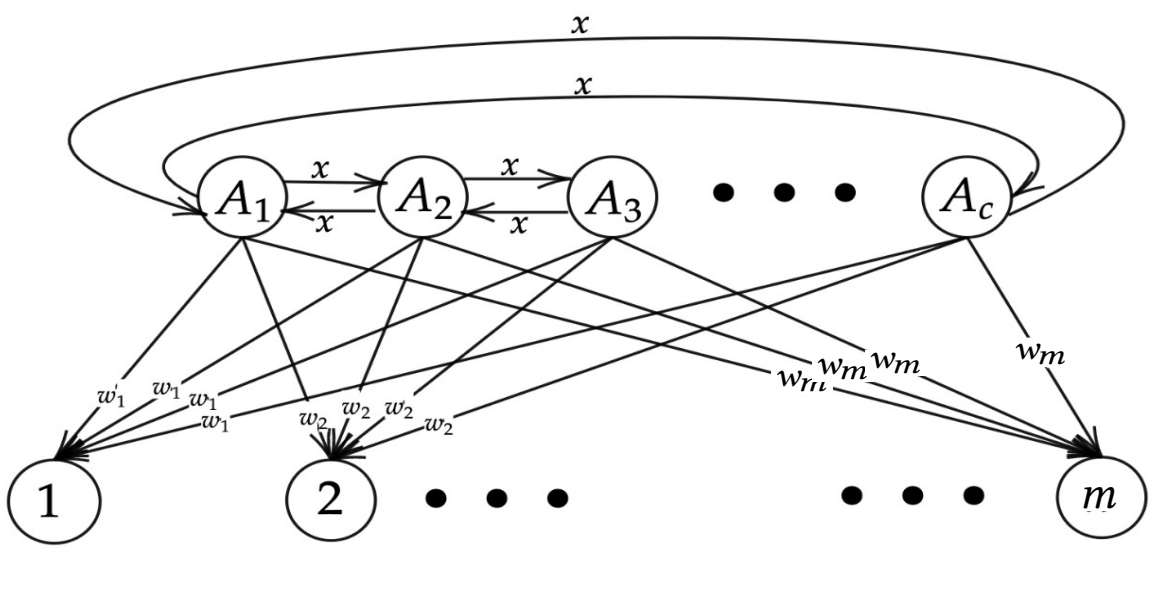
\begin{figure}[H]
\centering
\begingroup
\setlength{\unitlength}{1mm}
\setlength{\fboxsep}{0.5pt}
\begin{picture}(80,40)
\put(0,0){\includegraphics[height = 4cm, width=8cm]{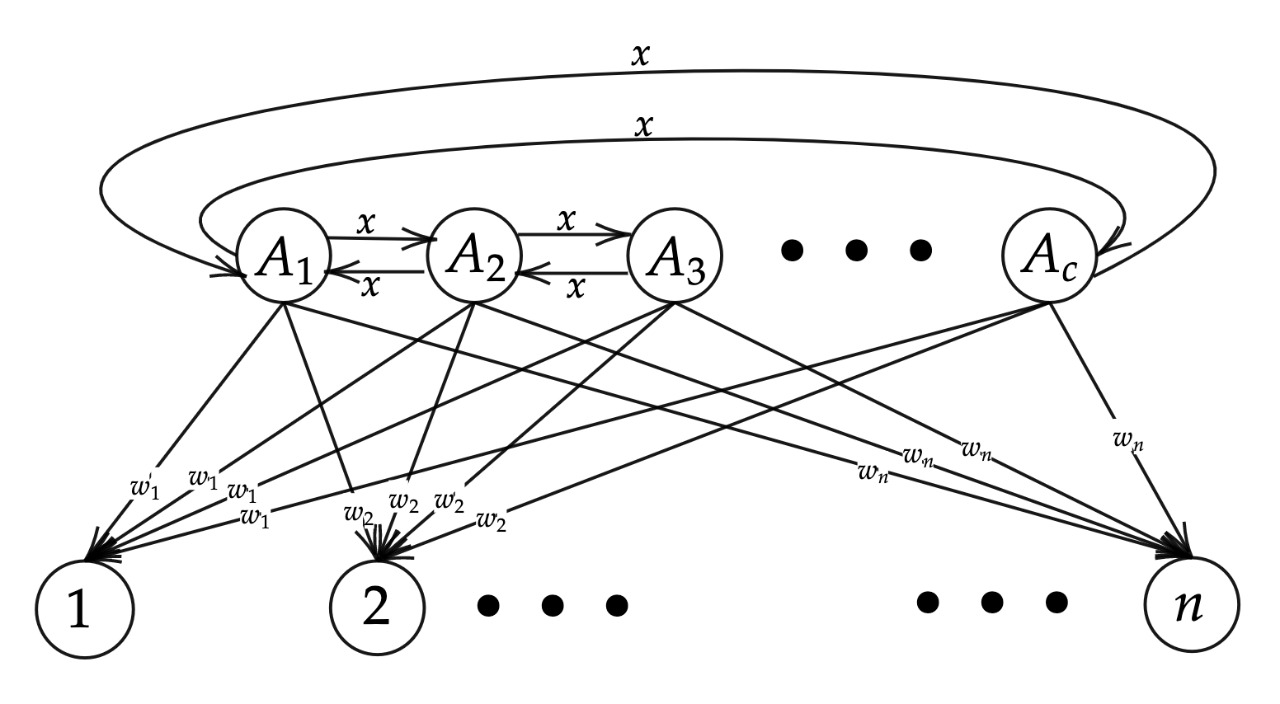}}
\put(54.7,13.3){\makebox(0,0){\colorbox{white}{$\scriptstyle w_m$}}}
\put(57.4,14.1){\makebox(0,0){\colorbox{white}{$\scriptstyle w_m$}}}
\put(60.9,14.5){\makebox(0,0){\colorbox{white}{$\scriptstyle w_m$}}}
\put(70.5,15.1){\makebox(0,0){\colorbox{white}{$\scriptstyle w_m$}}}
\put(74.4,5.6){\makebox(0,0){\colorbox{white}{\makebox[3mm]{$m$}}}}
\end{picture}
\endgroup
\caption{Gadget used for reduction}
\end{figure} 

 We will extend the reduction performed by ~\cite{olsen2008intractability} to \textsc{Target PageRank} Problem. We first define a general version of partition problem as follows: 

\begin{problem}
 \textbf{$(p_1, p_2, \dots , p_c)$}-\textit{Balanced Partition Problem : }
 We are given a vector $\boldsymbol{p}$ of $c\ge 1$ positive integers with $gcd(p_1,p_2, \dots, p_c) = 1$ and a multiset $\mathcal{W} = \{w_1, w_2, ..., w_m\}$ of $m$ positive integers. All integers are encoded in binary, and each occurrence of an item has its own index. Let $SUM(X)$ denote the sum of elements of multiset $X$. Does there exist a partition $\mathcal{W}_1, \mathcal{W}_2, \dots, \mathcal{W}_c$ of $\mathcal{W}$ such that $p_j \times SUM(\mathcal{W}_i) = p_i \times SUM(\mathcal{W}_j)$ and $p_j \times |\mathcal{W}_i| = p_i \times |\mathcal{W}_j|$ for all $i,j\in\{1,2,\dots,c\}$?
\end{problem}

 \begin{theorem}
 $(p_1, p_2, \dots, p_c)$-\textit{Balanced Partition Problem} is NP-complete, even when $c=2$ and $\boldsymbol{p}=(1,1)$.
 \end{theorem}

\begin{proof}
A certificate assigns each indexed occurrence to a block using $O(m\log(c+1))$ bits. Block sums, cardinalities, and the defining cross-products have polynomial bit length and can be computed and checked in polynomial time, proving membership in NP.

For NP-hardness, reduce from the NP-complete \textsc{Balanced Partition} problem~\cite[p.~404]{flake2004graph}, which requires two blocks of equal cardinality and equal weight sum. Retain every indexed item and its positive integer weight, including repeated values, and set $c=2$ and $\boldsymbol{p}=(1,1)$. Since $\gcd(1,1)=1$, this is a valid instance, whose constraints are exactly $SUM(\mathcal{W}_1)=SUM(\mathcal{W}_2)$ and $|\mathcal{W}_1|=|\mathcal{W}_2|$. The instances therefore have identical answers. This polynomial-time identity reduction proves NP-hardness for the stated restriction.
\end{proof}

 We prove $Theorem-1$ by reducing it from the \textbf{$(p_1, p_2, \dots, p_c)$}-\textit{Balanced Partition Problem} restricted to $c\ge 2$. The excluded case $c=1$ is trivial: the sole partition block contains every item.\\

We will use the weighted directed graph $G$ shown in Figure 12 as a gadget in our reduction. It has $m$ item nodes $N=\{1,2,\dots,m\}$ and $c$ center nodes $S=\{A_1,A_2,\dots,A_c\}$, so $n=|V|=m+c$. Let $x$ be a positive integer. For each $1\le i<c$, add a batch of $x$ copies of $A_i\rightarrow A_{i+1}$, and for each $1<i\le c$, add a batch of $x$ copies of $A_i\rightarrow A_{i-1}$. Add another batch of $x$ copies in each of the directions $A_1\rightarrow A_c$ and $A_c\rightarrow A_1$. These batches are additive: when $c=2$, they give $2x$ copies of $A_1\rightarrow A_2$ and $2x$ copies of $A_2\rightarrow A_1$. No self-loops are introduced. For every $A_i\in S$ and $j\in N$, include $w_j$ copies of $A_i\rightarrow j$. Write $W=\sum_{j=1}^{m}w_j$ for the total item weight, distinct from the multiset $\mathcal{W}$. Each center has out-degree $W+2x$, and every item is initially dangling. All multiplicities are stored as binary counts.\\
We claim that analogous lemmas will hold for $G$ as for the gadget used in ~\cite{olsen2008intractability}.
For any vertex subset $U$, write $M_U(E')=\sum_{v\in U}\boldsymbol{\tilde{\pi}}_v(E')$ for its total PageRank mass. The notation $\boldsymbol{\tilde{\pi}}_U(E')$ continues to denote the restricted vector, as in Section 3.

\begin{definition}
An addition multiset $E'$ of size $m$ is sum-optimal if for every $u\in N$, it contains an edge $(u,A_i)$ for some $A_i\in S$.
\end{definition}
Since $|N|=|E'|=m$, a sum-optimal multiset contains exactly one outgoing edge copy from each item to a center and no other additions. Thus its item-to-center groups partition $N$, and no dangling nodes remain after these additions.

\begin{lemma}
Consider an addition multiset $E'$ that is sum-optimal. Then,
\begin{equation}
\label{eq:reduction-sum-optimal}
M_S(E') = \frac{1 - (1-\alpha)\frac{m}{n}}{1+\alpha\frac{W}{W+2x}}
\end{equation}
\end{lemma}

\begin{proof}
Observe that the only nodes that link to a node in $N$ are in $S$, and all use a fraction $\frac{W}{W+2x}$ of their links on $N$. Also note that random teleportation contributes a page rank value of $\frac{(1-\alpha)}{n}$ to each node. Since the sum of page ranks of nodes in $N$ is $1-M_S(E')$, the following equation holds:
$$1-M_S(E')\ =\ (1-\alpha)\frac{m}{n} + \alpha M_S(E')
\frac{W}{W+2x}$$ Rearranging the terms gives the desired result.
\end{proof}

\begin{lemma}
Let $x$ satisfy the inequality $$2x > \frac{Wn^2}{m(1-\alpha)} - W$$ Let $E'$ be any sum-optimal addition multiset of size $m$ and let $L$ be any admissible addition multiset of size $m$ which is \textbf{not} sum-optimal. Repeated copies, including copies of existing edges, are allowed in $L$. Then the following holds: $$M_S(E') > M_S(L)$$
\end{lemma}

\begin{proof}
Since $L$ is not sum-optimal, some $u\in N$ has no outgoing edge to $S$. Teleportation contributes $(1-\alpha)/n$ to its PageRank. The existing edges from the centers to $u$ also carry positive mass, so $\boldsymbol{\tilde{\pi}}_u(L)>(1-\alpha)/n$. Consider the following cases for $u$:
\begin{itemize}
    \item $u$ is a sink: In this case, $u$ will spread $\alpha$ fraction of its page rank value equally among all nodes in the graph. Hence, 
    $$1-M_S(L) > (1-\alpha) \frac{m}{n} + \alpha \frac{1-\alpha}{n} \frac{m}{n}$$
    \item $u$ is not sink: In this case, $u$ will have outgoing edge(s) to nodes in $N$ only. So, $u$ will spread $\alpha$ fraction of its page rank among nodes in $N$. Hence, $$1-M_S(L) > (1-\alpha) \frac{m}{n} + \alpha \frac{1-\alpha}{n}$$
\end{itemize}
Combining the above inequalities, we have 
\begin{equation}
\label{eq:reduction-nonoptimal-mass}
    1-M_S(L) > (1-\alpha) \frac{m}{n} + \alpha \frac{1-\alpha}{n} \frac{m}{n}
\end{equation}
From \textit{Lemma 1} we have the following equality: 
\begin{equation}
\label{eq:reduction-optimal-mass}
    1-M_S(E')\ =\ (1-\alpha)\frac{m}{n} + \alpha M_S(E')\frac{W}{W+2x}
\end{equation}
Now observe that for $M_S(E') > M_S(L)$ to hold, we need $1-M_S(E') < 1-M_S(L)$. By using \eqref{eq:reduction-nonoptimal-mass} and \eqref{eq:reduction-optimal-mass} and the fact that $M_S(E')<1$, this holds whenever $\frac{(1-\alpha)m}{n^2} > \frac{W}{W+2x}$ which is true by the condition given in $Lemma \ 2$.
\end{proof}

\begin{lemma}
Let $E'$ be an arbitrary sum-optimal addition multiset, then define $A_{j_\leftarrow}$ = \{$i \in N$: $(i, A_j) \in E'$\} and $W_{A_{j_\leftarrow}} = \sum \limits_{i \in A_{j_\leftarrow}} w_i$. Let $p_1, p_2, \dots p_c$ be positive integers such that $gcd(p_1, p_2, \dots p_c)=1$ and suppose $x$ satisfies $$2x > max\{p_1, p_2, \dots, p_c\}\frac{\alpha Wn}{1-\alpha} - W$$
 Then the following 2 statements are equivalent:
\begin{enumerate}
    \item For all $i, j$ such that $1 \le i, j \le c,\ i \neq j$, we have $p_jW_{A_{i_\leftarrow}} = p_iW_{A_{j_\leftarrow}}$ and $p_j|A_{i_\leftarrow}|$ = $p_i|A_{j_\leftarrow}|$
    \item For all $i, j$ such that $1 \le i, j \le c, i \neq j$, we have $p_jM_{A_{i_\leftarrow}}(E') = p_i M_{A_{j_\leftarrow}}(E')$
\end{enumerate}
\end{lemma}

\begin{proof}
The center PageRank equations are given below. When $c=2$, the two neighbor terms in each equation refer to the same center and count the two additive batches; there are no interior-center equations.
\begin{equation}
\label{eq:reduction-first-center}
\begin{aligned}
\boldsymbol{\tilde{\pi}}_{A_1}(E')
&= \frac{1-\alpha}{n}
+ \alpha M_{A_{1_\leftarrow}}(E') \\
&\quad
+ \alpha
\left(
\boldsymbol{\tilde{\pi}}_{A_c}(E')
+ \boldsymbol{\tilde{\pi}}_{A_2}(E')
\right)
\frac{x}{W+2x}.
\end{aligned}
\end{equation}

\begin{equation}
\label{eq:reduction-last-center}
\begin{aligned}
\boldsymbol{\tilde{\pi}}_{A_c}(E')
&= \frac{1-\alpha}{n}
+ \alpha M_{A_{c_\leftarrow}}(E') \\
&\quad
+ \alpha
\left(
\boldsymbol{\tilde{\pi}}_{A_{c-1}}(E')
+ \boldsymbol{\tilde{\pi}}_{A_1}(E')
\right)
\frac{x}{W+2x}.
\end{aligned}
\end{equation}

For all $i$ such that $1<i<c$, we have:
\begin{equation}
\label{eq:reduction-interior-center}
\begin{aligned}
\boldsymbol{\tilde{\pi}}_{A_i}(E')
&= \frac{1-\alpha}{n}
+ \alpha M_{A_{i_\leftarrow}}(E') \\
&\quad
+ \alpha
\left(
\boldsymbol{\tilde{\pi}}_{A_{i-1}}(E')
+ \boldsymbol{\tilde{\pi}}_{A_{i+1}}(E')
\right)
\frac{x}{W+2x}.
\end{aligned}
\end{equation}

For all $i$ such that $1 \le i \le c$, we have:
\begin{equation}
\label{eq:reduction-group-mass}
    M_{A_{i_\leftarrow}}(E')\ =\ |A_{i_\leftarrow}| \frac{(1-\alpha)}{n} + \alpha M_S(E') \frac{W_{A_{i_\leftarrow}}}{W+2x}
\end{equation}\\\
\textbf{1 $\implies$ 2} : Assume that $p_jW_{A_{i_\leftarrow}} = p_i W_{A_{j_\leftarrow}}$ and $p_j|A_{i_\leftarrow}|$ = $p_i |A_{j_\leftarrow}|$ for a sum-optimal addition multiset $E'$. Then using \eqref{eq:reduction-group-mass} we get $p_jM_{A_{i_\leftarrow}}(E') = p_iM_{A_{j_\leftarrow}}(E')$. \\
\textbf{2 $\implies$ 1} : Using \eqref{eq:reduction-group-mass} and the fact that $p_jM_{A_{i_\leftarrow}}(E') = p_i M_{A_{j_\leftarrow}}(E')$ we get the following equation : \begin{equation}
\label{eq:reduction-balance}
 (p_j|A_{i_\leftarrow}| - p_i|A_{j_\leftarrow}|)\frac{1-\alpha}{n} =  (p_iW_{A_{j_\leftarrow}} - p_jW_{A_{i_\leftarrow}})\frac{\alpha M_S(E')}{W+2x}
\end{equation}
If $2x > max\{p_1, p_2, \dots p_c\}\frac{\alpha Wn}{1-\alpha} - W $ then we have
\begin{equation*}
\begin{aligned}
\frac{1-\alpha}{n}
&> \max\{p_1,p_2,\dots,p_c\}W\frac{\alpha}{W+2x}\\
&> |p_iW_{A_{j_\leftarrow}}-p_jW_{A_{i_\leftarrow}}|
\frac{\alpha M_S(E')}{W+2x}.
\end{aligned}
\end{equation*}
Now for equation \eqref{eq:reduction-balance} to hold we must have $p_j|A_{i_\leftarrow}| - p_i |A_{j_\leftarrow}| = 0$ which implies that $p_jW_{A_{i_\leftarrow}} = p_iW_{A_{j_\leftarrow}}$ and $p_j|A_{i_\leftarrow}|$ = $p_i |A_{j_\leftarrow}|$.\\
Hence the $Lemma\ 3$ holds.
\end{proof}

\begin{lemma}
For the network $G$ given in $Fig\ 12$ and sum-optimal addition multisets $E'$ of size $m$, there exists a bijective function $f_G$, depending only on $G$, which maps the group-mass vector $(M_{A_{i_\leftarrow}}(E'))_{i=1}^c$ to the center PageRank vector $\boldsymbol{\tilde{\pi}}_S(E')$.
\end{lemma}
\begin{proof}
Write $\boldsymbol{b}=(M_{A_{i_\leftarrow}}(E'))_{i=1}^c$ and $\boldsymbol{y}=\boldsymbol{\tilde{\pi}}_S(E')$, viewed as column vectors. Since $E'$ is sum-optimal, it changes no outgoing edges of the centers and leaves no dangling nodes. Define the matrix $R\in\mathbb{R}^{c\times c}$ by
\begin{equation*}
R_{ij}=\frac{w_{A_jA_i}}{W+2x},\qquad 1\le i,j\le c,
\end{equation*}
where $w_{A_jA_i}$ is the multiplicity in $G$. Thus column $j$ records the edge-following probabilities from $A_j$ to the centers, before damping. Each column sums to $2x/(W+2x)$. For $c=2$, the additive batches give $R_{12}=R_{21}=2x/(W+2x)$ and $R_{11}=R_{22}=0$.

Let $\tau=(1-\alpha)/n$, let $\boldsymbol{1}\in\mathbb{R}^c$ be the all-ones vector, and let $I$ be the $c\times c$ identity matrix. Equations~\eqref{eq:reduction-first-center}--\eqref{eq:reduction-interior-center} are exactly
\begin{equation*}
\boldsymbol{y}=\tau\boldsymbol{1}+\alpha\boldsymbol{b}+\alpha R\boldsymbol{y},
\qquad
(I-\alpha R)\boldsymbol{y}=\tau\boldsymbol{1}+\alpha\boldsymbol{b}.
\end{equation*}
Because $R$ is nonnegative, its induced matrix $1$-norm satisfies
\begin{equation*}
\|\alpha R\|_1=\max_j\sum_{i=1}^c|\alpha R_{ij}|
=\frac{2\alpha x}{W+2x}<1.
\end{equation*}
If $(I-\alpha R)\boldsymbol{z}=0$, then $\boldsymbol{z}=\alpha R\boldsymbol{z}$, so
\begin{equation*}
\|\boldsymbol{z}\|_1\le\|\alpha R\|_1\|\boldsymbol{z}\|_1.
\end{equation*}
The strict bound forces $\boldsymbol{z}=0$. Hence the square matrix $I-\alpha R$ is invertible, and the center equations have a unique solution for every $\boldsymbol{b}$.

Since $\alpha>0$, the affine map $f_G:\mathbb{R}^c\to\mathbb{R}^c$ and its inverse are
\begin{equation*}
\begin{aligned}
f_G(\boldsymbol{b})&=(I-\alpha R)^{-1}(\tau\boldsymbol{1}+\alpha\boldsymbol{b}),\\
f_G^{-1}(\boldsymbol{y})&=\alpha^{-1}\bigl((I-\alpha R)\boldsymbol{y}-\tau\boldsymbol{1}\bigr).
\end{aligned}
\end{equation*}
For fixed $\alpha$, their coefficients depend only on $G$, not on $E'$. Restricting $f_G$ to the group-mass vectors realized by sum-optimal additions gives the claimed bijection onto the corresponding center PageRank vectors.
\end{proof}

We are now ready to prove $Theorem\ 1$.

\begin{proof}
Consider an instance of $(p_1, p_2, \dots p_c)$-Balanced Partition with $c\ge 2$ and a nonempty item multiset $\mathcal{W} = \{w_1, w_2, ..., w_m\}$, so $m\ge 1$. Let $\ell_{\mathrm{src}}$ be the binary input length and set $p_{\max}=\max_i p_i$. Choose the gadget parameter explicitly as
\begin{equation*}
\begin{aligned}
H&=\max\left\{0,\frac{p_{\max}\alpha Wn}{1-\alpha}-W,
\frac{Wn^2}{m(1-\alpha)}-W\right\},\\
x&=1+\left\lfloor\frac{H}{2}\right\rfloor.
\end{aligned}
\end{equation*}
Then $x$ is a positive integer and $2x>H$, so both strict bounds in Lemmas 2 and 3 hold. Since $\alpha$ is fixed and rational, $H$ and $x$ can be computed exactly in polynomial bit time, and
\begin{equation*}
\log_2(x+1)=O\bigl(\log_2(p_{\max}+1)+\log_2(W+1)+\log_2(n+1)\bigr).
\end{equation*}
The explicit item and parameter lists give $m,c\le\ell_{\mathrm{src}}$, and $W$ also has polynomially many bits. Construct $G$ as above with $n=m+c$ vertices and at most $mc+2c$ nonzero ordered-pair multiplicities. Store one binary count per such pair, including the count $2x$ for each center-to-center direction when $c=2$; do not enumerate edge copies. Thus $G$, the target set $T=S$, and the budget $B=m$ have polynomial encoding size and are constructible in polynomial bit time.

To define the targets, let $\sigma$ be the right-hand side of~\eqref{eq:reduction-sum-optimal} and set $p_{\Sigma}=\sum_{i=1}^c p_i$. The desired group masses have total $1-\sigma$ and satisfy the proportionality conditions in statement 2 of Lemma 3, so they are
\begin{equation*}
b_i^*=(1-\sigma)\frac{p_i}{p_{\Sigma}},\qquad 1\le i\le c.
\end{equation*}
Using $R$ and $\tau$ from Lemma 4, define the target score vector by
\begin{equation*}
\boldsymbol{t}=f_G(\boldsymbol{b}^*),\qquad
(I-\alpha R)\boldsymbol{t}=\tau\boldsymbol{1}+\alpha\boldsymbol{b}^*.
\end{equation*}
The numerator defining $\sigma$ lies in $(0,1)$ and its denominator exceeds one, so $0<\sigma<1$ and $b_i^*>0$. Set $q=2x/(W+2x)$. Since $R\ge0$ and $\|\alpha R\|_1=\alpha q<1$, the convergent series $(I-\alpha R)^{-1}=\sum_{k=0}^{\infty}(\alpha R)^k$ is entrywise nonnegative. Thus $\boldsymbol{t}\ge\tau\boldsymbol{1}+\alpha\boldsymbol{b}^*>\boldsymbol{0}$ entrywise. Moreover, $\boldsymbol{1}^{\top}R=q\boldsymbol{1}^{\top}$. Writing $s=\sum_{i=1}^c\boldsymbol{t}_{A_i}$ and summing the target equations gives
\begin{equation*}
\begin{aligned}
(1-\alpha q)s
&=c\tau+\alpha(1-\sigma)\\
&=1-m\tau-\alpha\sigma\\
&=(1-\alpha q)\sigma,
\end{aligned}
\end{equation*}
where $n=m+c$ and the definition of $\sigma$ gives $(1+\alpha(1-q))\sigma=1-m\tau$. Since $1-\alpha q>0$, we obtain $s=\sigma$ and $0<\boldsymbol{t}_{A_i}\le\sigma<1$ for every $i$. The targets are therefore valid absolute scores on $S$, without renormalization, even when no balanced partition exists.

The numerators and denominators of $\sigma$, $\boldsymbol{b}^*$, and all coefficients of this $c\times c$ nonsingular system have polynomial bit length in $\ell_{\mathrm{src}}$. Multiplying the system by the product of its coefficient and right-hand-side denominators yields an integer augmented system with polynomial-bit entries. In general, for a $k\times k$ integer system whose augmented entries have at most $h$ bits, every minor has magnitude at most $k!2^{kh}$ and therefore $O(k(h+\log k))$ bits. Cramer's rule expresses each solution entry as a ratio of two such determinants. Each determinant can be computed by fraction-free Gaussian elimination with pivoting using polynomially many integer operations: its stored entries are minors, and its temporary products and exact divisions also involve only polynomial-bit integers. Consequently, the exact rational targets have polynomial encoding length and can be computed in polynomial bit time. This establishes the required construction-size and arithmetic bounds.

Now we will show that there exists a $(p_1, p_2, \dots p_c)$-balanced partition of $\mathcal{W}$ if and only if there exists a solution $E'$ to the \textsc{Target PageRank} instance.\\  \\
1) Suppose $\mathcal{W}_1,\mathcal{W}_2,\dots,\mathcal{W}_c$ form such a balanced partition. Track individual occurrences by partitioning their indices as $N=I_1\mathbin{\dot\cup}\cdots\mathbin{\dot\cup}I_c$, where $\mathcal{W}_j=\{\!\{w_i:i\in I_j\}\!\}$. Define the addition multiset
\begin{equation*}
E'=\biguplus_{j=1}^{c}\ \biguplus_{i\in I_j}\{(i,A_j)\}.
\end{equation*}
Each indexed item contributes exactly one edge, so $|E'|=m=B$, no self-loops are added, and $E'$ is sum-optimal. Equal-valued occurrences remain distinct. By Lemmas 3 and 4, $\boldsymbol{\tilde{\pi}}_{A_j}(E')=\boldsymbol{t}_{A_j}$ for every $j\in\{1,\dots,c\}$. Hence $E'$ is a solution of the \textsc{Target PageRank} instance. \\
2) Suppose there exists a solution $E'$ to the \textsc{Target PageRank} instance. Since $T=S$, attaining the targets gives $M_S(E')=\sum_{i=1}^c\boldsymbol{t}_{A_i}=\sigma$. A sum-optimal multiset always exists, for example by adding $(u,A_1)$ for every $u\in N$. By Lemmas 1 and 2, every non-sum-optimal addition multiset of size $m$ has center mass strictly below $\sigma$. Hence $E'$ is sum-optimal. Its groups $A_{i_\leftarrow}$, $1\le i\le c$, therefore partition the item indices $N$. Lemmas 3 and 4 give, for all $1\le i,j\le c$,
\begin{equation*}
\begin{aligned}
p_jW_{A_{i_\leftarrow}}&=p_iW_{A_{j_\leftarrow}},\\
p_j|A_{i_\leftarrow}|&=p_i|A_{j_\leftarrow}|.
\end{aligned}
\end{equation*}
Define
\begin{equation*}
\mathcal{W}_i=\{\!\{w_j:j\in A_{i_\leftarrow}\}\!\},\qquad 1\le i\le c.
\end{equation*}
These are multisets of item occurrences, not their numerical sums. They partition $\mathcal{W}$, and $SUM(\mathcal{W}_i)=W_{A_{i_\leftarrow}}$ and $|\mathcal{W}_i|=|A_{i_\leftarrow}|$. Thus they form the required balanced partition.\\

It remains to establish membership in NP in the binary encoding of Section 3. Let $\ell_{\mathrm{in}}$ be the input length of an arbitrary instance $(G,T,\boldsymbol{t},B)$. Since the vertex list is explicit, $n\le\ell_{\mathrm{in}}$. A certificate consists of the counts $x_{uv}$ for the $n(n-1)$ ordered pairs with $u\ne v$, in a fixed order induced by the vertex list, with $x_{uu}=0$ implicit. Every feasible count is at most $B$, so the certificate uses $O(n^2(1+\log_2(B+1)))$ bits, including when $B=0$, and is polynomial in $\ell_{\mathrm{in}}$. The verifier checks that the counts are nonnegative integers, that each is at most $B$, and that their sum is exactly $B$.

Without expanding edge copies, form $w'_{uv}=w_{uv}+x_{uv}$ and $d'_u=\sum_v w'_{uv}$. These integers have $O(\ell_{\mathrm{in}}+\log n)$ bits. Recompute the edge-following transition matrix $Q$, with destinations indexing rows and sources indexing columns, by
\begin{equation*}
Q_{vu}=\begin{cases}
w'_{uv}/d'_u,&d'_u>0,\\
1/n,&d'_u=0.
\end{cases}
\end{equation*}
Thus a formerly dangling node uses only its actual outgoing edges once $d'_u>0$. Every column of $Q$ sums to one, so $\|\alpha Q\|_1=\alpha<1$. The same kernel argument as in Lemma 4 shows that $I-\alpha Q$ is invertible. The verifier computes $\boldsymbol{\tilde{\pi}}(E')$ as the exact rational solution of
\begin{equation*}
(I-\alpha Q)\boldsymbol{\tilde{\pi}}(E')
=\frac{1-\alpha}{n}\boldsymbol{1},
\end{equation*}
where the identity and all-ones vector now have dimension $n$.

For an explicit arithmetic bound, write the fixed damping factor as $\alpha=r_\alpha/s_\alpha$ with positive integers $r_\alpha<s_\alpha$. Multiplying this system by
\begin{equation*}
D=s_\alpha n\prod_{u:\,d'_u>0}d'_u
\end{equation*}
clears every denominator, including dangling-node and teleportation terms. The empty product is one. The integer $D$ has $O(n(\ell_{\mathrm{in}}+\log n))$ bits, and the resulting integer augmented entries have the same bit bound. Applying the determinant and fraction-free elimination bounds above with $k=n$ bounds every solution numerator and denominator by $O(n^2(\ell_{\mathrm{in}}+\log n))$ bits and permits computation in polynomial bit time. Finally, the verifier checks nonnegativity and normalization as in Definition~\ref{def:pagerank}, and compares $\boldsymbol{\tilde{\pi}}_v(E')$ with the input fraction $\boldsymbol{t}_v=a_v/b_v$ for each $v\in T$ by integer cross-multiplication. All comparisons are exact and have polynomial bit cost; no numerical tolerance is used. Every yes instance has a polynomial-size certificate accepted by this verifier, and acceptance guarantees a feasible solution. This proves membership in NP.
\end{proof}

\begin{corollary}
\label{cor:at-most-target-pagerank}
The variant of Problem~1 allowing an addition multiset $E'$ with $|E'|\le B$, with all other conventions unchanged, is NP-complete even when $|T|=2$.
\end{corollary}

\begin{proof}
Use the construction of Theorem~1 with $c=2$, $\boldsymbol{p}=(1,1)$, and $B=m$. Let $\tau=(1-\alpha)/n$ and $r=W/(W+2x)$. The chosen $x$ satisfies $r<\tau m/n$, while Lemma~1 gives $1-\sigma=m\tau+\alpha\sigma r$ with $0<\sigma<1$.

If an addition multiset $L$ has $|L|<m$, some item $u\in N$ remains dangling, since each added edge has only one source. Its PageRank is at least $\tau$, and its edge-following transition assigns probability $m/n$ to $N$. Summing PageRank over $N$ and retaining only teleportation and this dangling-node contribution yields
\begin{equation*}
\begin{aligned}
1-M_S(L)&\ge m\tau+\alpha\frac{m}{n}\boldsymbol{\tilde{\pi}}_u(L)\\
&\ge m\tau+\alpha\tau\frac{m}{n}\\
&>m\tau+\alpha\sigma r=1-\sigma.
\end{aligned}
\end{equation*}
Thus $M_S(L)<\sigma=\sum_{i=1}^{c}\boldsymbol{t}_{A_i}$, so fewer than $m$ additions cannot attain the targets. Exact-budget solutions are still allowed, and every solution with at most $B=m$ additions must use exactly $m$. The same polynomial-time reduction therefore proves NP-hardness.

Membership in NP uses the binary count certificate and exact rational verifier from Theorem~1, replacing only the budget check $\sum_{u,v}x_{uv}=B$ by $\sum_{u,v}x_{uv}\le B$. Each count is still at most $B$, so the certificate-size and arithmetic bounds are unchanged.
\end{proof}

\balance
\section{Proof of Theorem 2}

\begin{proof}
Fix a finite constant $\rho\ge1$ and suppose that a deterministic polynomial-time multiplicative $\rho$-approximation algorithm exists. Start with an instance of the NP-complete restriction of Balanced Partition with $c=2$ and $\boldsymbol{p}=(1,1)$ used in Theorem~1. An empty item multiset is trivially a yes-instance and can be handled directly. Otherwise, apply the polynomial-time construction in Theorem~1 to obtain $I=(G,T,\boldsymbol{t},B)$ with $m\ge1$ items, $n=m+2$ vertices, and $B=m$.

The optimization instance $I$ is always feasible: adding one edge $(u,A_1)$ for each item $u\in N$ uses exactly $B$ edges and introduces no self-loops, whether or not the source instance has a balanced partition. There are finitely many feasible addition count vectors, so the minimum $\operatorname{OPT}(I)$ defined in Section~3 is attained. The equivalence established in Theorem~1 gives
\begin{equation*}
\operatorname{OPT}(I)=0
\quad\Longleftrightarrow\quad
\text{the source instance is a yes-instance}.
\end{equation*}

Run the assumed approximation algorithm on $I$. Its output is an addition multiset $E'$ represented by binary-encoded counts satisfying
\begin{equation*}
x_{uv}\in\mathbb{Z}_{\ge0},\qquad x_{uu}=0,\qquad
\sum_{u,v\in V}x_{uv}=B.
\end{equation*}
Feasibility, nonnegativity of the objective, and the multiplicative guarantee, including at zero optimum, imply
\begin{equation*}
0\le\operatorname{OPT}(I)\le F_I(E')\le\rho\operatorname{OPT}(I).
\end{equation*}
Thus $F_I(E')=0$ if and only if $\operatorname{OPT}(I)=0$.

This zero test is computable in polynomial bit time. As shown in the NP-membership argument for Theorem~1, the counts use $O(n^2(1+\log_2(B+1)))$ bits. Without expanding edge copies, add the multiplicities $w_{uv}+x_{uv}$ and recompute all out-degrees and dangling-node transitions as in Section~3. For the fixed rational damping factor, the same verifier solves the PageRank system exactly with polynomial-bit rational arithmetic and checks $\boldsymbol{\tilde{\pi}}_v(E')=\boldsymbol{t}_v$ for every $v\in T$ by integer cross-multiplication. These equalities hold exactly when $F_I(E')=0$; no numerical tolerance is used.

Accept the source instance if and only if this exact zero test succeeds. The construction, approximation algorithm, and verification all run in polynomial bit time, so this decides the NP-complete source problem in polynomial time. It follows that $P=NP$, proving the stated conditional inapproximability result.
\end{proof}

\bibliographystyle{unsrt}
\bibliography{references}

\end{document}